\documentclass[apj]{emulateapj}
\usepackage[colorlinks,linkcolor={blue},citecolor={blue},urlcolor={red}]{hyperref}
\usepackage{epsfig,graphicx,natbib,amsmath,amsfonts,amssymb}
\usepackage{color}
\usepackage{multirow}
\usepackage{booktabs}
\usepackage{amssymb}
\usepackage{threeparttablex} 
\usepackage{booktabs} 

\usepackage{longtable}
\usepackage{array} 
\usepackage[dvipsnames]{xcolor}  

\newcommand{\re}{\Reff}
\newcommand{\msolar}{${\rm M}_\odot$}
\newcommand{\mstar}{$M_\ast$}
\newcommand{\lgmstar}{$\log_{10}$($M_\ast/h^{-2}$\msolar)}

\newcommand{\Reff}{{$R_{\rm e}$}}

\newcommand{\msun}{M$_{\odot}$}

\newcommand{\sersic}{S\'{e}rsic}

\newcommand{\myemail}{\email{ecwang16@ustc.edu.cn(EW), whywang@ustc.edu.cn(HW)}}

\defcitealias{Wang-18a}{Paper I}
\defcitealias{Wang-18b}{Paper II}

\shorttitle{The dearth of difference between central and satellite galaxies}
\shortauthors{Wang et al.}

\graphicspath{{fig/}}

\begin{document}

\title{The dearth of difference between central and satellite galaxies III. Environmental dependencies of mass-size and mass-structure relations} 
\author{
Enci Wang\altaffilmark{1,2,3},
Huiyuan Wang\altaffilmark{2,3},
Houjun Mo\altaffilmark{4,5}, 
Frank C. van den Bosch\altaffilmark{6},
Xiaohu Yang\altaffilmark{7,8}
} \myemail

\altaffiltext{1}{Department of Physics, ETH Zurich, Wolfgang-Pauli-Strasse 27, CH-8093 Zurich, Switzerland}
\altaffiltext{2}{CAS Key Laboratory for Research in Galaxies and Cosmology, Department of Astronomy, University of Science and Technology of China, Hefei 230026, China}
\altaffiltext{3}{School of Astronomy and Space Science, University of Science and Technology of China, Hefei 230026, China}
\altaffiltext{4}{Department of Astronomy, University of Massachusetts, Amherst MA 01003-9305, USA}
\altaffiltext{5}{Tsinghua Center of Astrophysics \& Department of Physics, Tsinghua University, Beijing 100084, China}

\altaffiltext{6}{Department of Astronomy, Yale University, P.O. Box 208101, New Haven, CT 06520-8101, USA}
\altaffiltext{7}{Department of Astronomy, Shanghai Jiao Tong University, Shanghai 200240, China}
\altaffiltext{8}{IFSA Collaborative Innovation Center, Shanghai Jiao Tong University, Shanghai 200240, China}

\begin{abstract}  
  As demonstrated in \citetalias{Wang-18a}, the quenching properties of central
  and satellite galaxies are quite similar as long as both stellar
  mass and halo mass are controlled. Here we extend the
  analysis to the size and
  bulge-to-total light ratio (B/T) of galaxies. In general
  central galaxies have size-stellar mass and B/T-stellar mass
  relations different from satellites. However, the differences are
  eliminated when halo mass is controlled. We also
  study the dependence of size and B/T on halo-centric distance and
  find a transitional stellar mass ($M_{\rm *,t}$) at given halo mass
  ($M_h$), which is about one fifth of the mass of the central 
  galaxies in halos of mass $M_h$. The transitional stellar masses
  for size, B/T and quenched fraction are similar over the whole halo
  mass range, suggesting a connection between the
  quenching of star formation and the structural evolution of galaxies.  
  Our analysis further suggests that the classification
  based on the transitional stellar mass is more fundamental than the
  central-satellite dichotomy, and provide a more reliable way to 
  understand the environmental effects on galaxy properties. 
  We compare the observational
  results with the hydro-dynamical simulation, EAGLE and the
  semi-analytic model, L-GALAXIES. The EAGLE simulation successfully
  reproduces the similarities of size for centrals and satellites and
  even $M_{\rm *,t}$, while L-GALAXIES fails to recover the
  observational results.
\end{abstract}


\section{Introduction}
\label{sec:introduction}

Thanks to the large photometric and spectroscopic surveys of galaxies
at both low and high redshift \citep[e.g.][]{York-00, Lilly-07,
  Abazajian-09, Grogin-11}, our understanding of the formation and
evolution of galaxies is rapidly enriched.  One of the most remarkable
findings is the strong size evolution of galaxies since redshift of
about 2. In particular, while passive galaxies are observed to be 3-5 times
smaller in size at redshift greater than 1 with respect to their
counterparts of similar stellar mass in the local Universe
\citep{Shen-03, Daddi-05, Trujillo-06, Toft-07, Cimatti-08,
  vanderWel-08, vanDokkum-08, Buitrago-08, Damjanov-11, Raichoor-12,
  Mei-12,Cassata-13, vanderWel-14}, star-forming
galaxies are only about two times smaller in size than their
counterparts today.  This suggests an entangled evolution of size (or
structural properties) and star formation activities of galaxies.
Although the main driver of the size evolution of galaxies is
  still under debate, several suggestions have been made to account
  for these observational findings. A number of studies have argued that
  whereas active galaxies grow in size due to in-situ star formation,
  passive galaxies mainly grow in size due to frequent, minor mergers
  \citep{Shen-03,Khochfar-Silk-06, Naab-Khochfar-Burkert-06, Oser-10}.
  However, others have argued that larger passive galaxies simply
  have quenched more recently, an effect known as ``progenitor bias''
  \citep{Newman-12, Carollo-13a, Poggianti-13, Fagioli-16}. A third
  possibility is that the size evolution is driven by a quasi-adiabatic
  expansion, which results from the ejection of large amounts of mass
  from galaxies via, e.g., quasar feedback \citep{Fan-08, Damjanov-09,
    Ragone-Figueroa-Granato-11}.
         
In the standard cosmological model of structure formation, dark matter
halos assemble hierarchically through mergers of smaller halos, and
galaxies are assumed to form in these halos.  A natural consequence of
this process is that the properties of galaxies can be significantly
affected by the formation histories of their host halos. Indeed,
galaxies in massive halos are believed to suffer from a series of
environmental effects, such as major or minor mergers
\citep[e.g.][]{Conselice-Chapman-Windhorst-03, Cox-06, Cheung-12,
  Peng-14}, tidal interaction \citep[e.g.][]{Gunn-Gott-72, Read-06},
ram-pressure stripping \citep[e.g.][]{Abadi-Moore-Bower-99, Wang-15,
  Poggianti-17} or strangulation \citep{Larson-80, Balogh-Navarro-Morris-00,
  vandenBosch-08}, and galaxy harassment
\citep{Farouki-Shapiro-82, Moore-96}.  These environmental effects may
both change the size/structure of a galaxy and prevent its growth by
quenching its star formation.  Major merger is an effective way to
turn gas-rich disk galaxies into passive ellipticals, as shown in
numerical simulations \citep{Toomre-Toomre-72, Farouki-Shapiro-82,
  Negroponte-White-83}, while minor mergers may cause size growth of
massive early-type galaxies 
\citep[e.g.][]{DeLucia-06, DeLucia-11, Fontanot-11,
  Khochfar-11, Shankar-13, Wilman-13}.  Strong tidal stripping can
effectively strip stars in the outer regions of galaxies
\citep[e.g.][]{Read-06}, while galaxy harassment can disturb galaxies
and cause enhancements of star formation in the central regions,
making the galaxies to be more compact \citep{Fujita-98}.
Ram-pressure stripping and strangulation can remove cold and hot gas
for star formation, and inhibit the growth of galaxy sizes by
suppressing star formation \citep{vandenBosch-08,
  Peng-Maiolino-Cochrane-15, Quilis-Planelles-Ricciardelli-17}.  Some
galaxy formation models have invoked parts of these mechanisms and
predicted significant dependence of galaxy size on environments
\citep[see e.g.][]{Guo-09, Shankar-13, Shankar-14}.

In many contemporary galaxy formation models, central and satellite
galaxies are assumed to experience different environmental
processes. Therefore, attempts have been made to constrain these
processes by comparing these two populations.  By using the SDSS
galaxy group catalog of \cite{Yang-07}, \cite{Weinmann-09}
found that late-type satellite galaxies have smaller radii and larger
concentrations than late-type central galaxies of the same stellar
mass, but no difference is found for early-type galaxies. In contrast,
\cite{Lim-17} found a significant difference between centrals and
satellites for both late- and early-type
galaxies. \cite{Spindler-Wake-17} found that the mass-size relation
for centrals and satellites shows no significant differences. However,
when fixing the velocity dispersion, they found significant
differences in size and stellar mass between centrals and satellites
in the passive population.  More recently, \cite{Bluck-19} found that
the bulge to total ratio (B/T) of satellites is larger than that of
centrals at given stellar mass, and the relation between B/T and halo
mass or local over-density for centrals is stronger than that for
satellites.

However, there is growing evidence that centrals and satellites may
not be as different as usually assumed, in particular with
respect to their quenching properties \citep{Hirschmann-14,
  Knobel-15, Wang-18a, Wang-18b, Wang-18c}. For example, 
\cite{Wang-18a}, hereafter \citetalias{Wang-18a}, presented a 
comprehensive comparison of the two
populations and found that centrals and satellites show similar
quenching properties and a similar prevalence of AGN activity,
  when the comparison is made at given both stellar and halo mass.  In
contrast, using the GAMA group catalog of \citep{Robotham-11},
\cite{Davies-19} found clear differences in the fractions of passive
galaxies between centrals and satellites, even if both stellar and
halo mass are controlled. They also found that the
quenched fraction of central galaxies is independent of, or even
decreases slightly with, the host halo mass, which is different from
the results obtained previously, both from observations and from
theoretical models \citep[e.g.][]{Hirschmann-14, Henriques-17,
  Wang-18a, Wang-18b}. It is unclear what causes these
  discrepancies. Part of it may be due to the differences in the
  group finders used, which are known to suffer from errors in
  central/satellite identification and in the assignment of halo
  mass. Indeed, as shown in \cite{Robotham-11}, their group finder
  sometimes identifies small galaxies (less than $10^{10.2}$\msun) as
  central galaxies of massive galaxy clusters (more massive than
  $10^{14}$\msun), in conflict with the central galaxy mass - halo
  mass relation revealed by various methods
  \citep[e.g.][]{Yang-Mo-vandenBosch-03, Yang-07, Guo-10,
    Leauthaud-12, Moster-Naab-White-13, Wang-13, Hudson-15, Han-15}.
  The group catalog of \cite{Yang-07} used by \cite{Wang-18a}
  is not free of errors either \citep[e.g.,][]{Campbell-15}.

 Given these discrepancies and the potential errors resulting
  from the use of imperfect galaxy group finders, it is prudent that
  any results be interpreted with great care. In \cite{Wang-18b},
  hereafter \citetalias{Wang-18b}, we therefore used a forward modeling to compare
  the findings of \citetalias{Wang-18a} with two galaxy formation models; the
  hydrodynamical EAGLE\footnote{Evolution and Assembly of GaLaxies and
    their Environments} simulation \citep[][]{Schaye-15, Crain-15} and
  the L-GALAXIES semi-analytical model \citep{Henriques-15,
    Henriques-17}. In particular, the observational data was compared
  to mock samples that were constructed from EAGLE and L-GALAXIES and  
  analyzed using the same galaxy group finder as used for the SDSS
  data. This assures that the model-data comparison is fair, and also
  allows one to check how imperfections in the group finder impact the
  results.  This analysis showed that L-GALAXIES fails to match the
  observed trends, while the EAGLE simulation nicely reproduces the
  observed similarities between centrals and satellites found in Paper
  I. The analysis presented in \citetalias{Wang-18b} also revealed, though, that
  the results based on the \cite{Yang-07} group finder can cause
  centrals and satellites to appear more similar than they really
  are. This reiterates the conclusion of \cite{Campbell-15}, that a
  proper interpretation of statistics inferred from a galaxy group
  catalog is best achieved using forward modeling (i.e. running group
  finders over mock data).

 In this paper we extend the analyses in \citetalias{Wang-18a} and \citetalias{Wang-18b} 
 by examining to what extent centrals and satellites are different with
  regard to their sizes and structural properties, especially when
  both are controlled for both stellar and halo mass. As in \citetalias{Wang-18b},
  we will compare the results obtained from the SDSS galaxy group
  catalog of \cite{Yang-07} to mock data extracted from the EAGLE
  simulation and the L-GALAXIES semi-analytical model. This will shed
  light on whether current galaxy formation models can reproduce the
  sizes and structural properties of galaxies as function of
  environment. The two environmental indicators to be considered in
  this study are host halo mass (i.e., the assigned group mass), and
  halo-centric distance (i.e., the projected distance from the
  luminosity-weighted group center).
  
This paper is organized as follows. In Section \ref{sec:data},
we present the observational galaxy sample and group catalog, as well
as samples generated from the simulation and semi-analytical
model. In Section \ref{sec:results}, we study the size and structural
properties of galaxies hosted by different halos and at different
locations within halos in the observational data. In Section
\ref{sec:model}, we compare the observational results with galaxy
formation models. Finally, we summarize our results and discuss their
implications in Section \ref{sec:summary}.

\section{Data}
\label{sec:data}

\subsection{Galaxies and groups from SDSS}
\label{subsec:2.1}

The observational data, such as galaxy and group catalogs, used in
this work are the same as in \citetalias{Wang-18a}. Here we only briefly describe
the sample selection and parameter measurements.  The reader is
referred to \citetalias{Wang-18a} for details.

The galaxy sample is originally selected from the NYU-VAGC
\citep[][]{Blanton-05a} of the Sloan Digital Sky Survey (SDSS) DR7
\citep{Abazajian-09}. Galaxies are selected to have 1) redshift in the
range of $0.01<z<0.2$, 2) spectroscopic completeness $C>0.7$ and 3)
magnitude limit $r=17.72$ mag.  The first two criteria ensure that the
sample galaxies are the same as those in the group catalog constructed
by \cite{Yang-07}, which provides the basic
environmental information, such as central-satellite classification
and halo mass ($M_{\rm h}$).

The stellar masses used here, taken from the group catalog of 
\cite{Yang-07}, are computed using the empirical relation between 
the stellar mass-to-light ratio and the $g-r$ colour as given 
in \cite{Bell-03}, adopting a \cite{Kroupa-01} initial mass function. 
The overall uncertainty in this stellar mass estimate is about   
0.15 dex \citep{Bell-03}. The magnitude and colour of galaxies are 
taken from the NYU-VAGC \citep{Blanton-05a}, which is based on SDSS DR4 
\citep{Adelman-McCarthy-06}, but includes a set of significant 
improvements over the original pipelines. The systematic calibration 
error in the photometry is about $1$-$2\%$ across the sky.

In \cite{Yang-07}, the most massive galaxy in a group/cluster is defined as the central galaxy, and the others are satellites. 
The halo masses of these galaxy groups/clusters are estimated by
abundance matching galaxy groups, rank-ordered by the total stellar mass 
of all member galaxies with $r$-band absolute magnitude brighter 
than $-19.5$ mag, to dark matter halos rank-ordered by halo mass.   
Dark matter halos are defined to have a mean over-density of 180, 
and the halo mass is defined to be the mass of dark matter 
enclosed by the radius within which the mean over-density is 180. 
The typical uncertainty of halo mass is 0.25 dex \citep{Yang-07}.
The center of a group is defined as the luminosity-weighted center of 
member galaxies.  Thus, central galaxies are not always located at the 
centers of the groups. For each galaxy, we defined a scaled halo-centric 
radius ($R_{\rm p}/r_{\rm 180}$), which is the projected distance 
from the galaxy to the host group center scaled by the virial 
radius of the host halo \citep{Yang-07}.


In this paper, we investigate two structural properties, 
the size and morphology, of SDSS galaxies. We adopt the SDSS 
$r$-band half-light radius (\re) as the size of a galaxy, and 
the $r$-band bulge-to-light ratio (B/T) to represent the morphology. 
The two parameters are taken from the UPenn photometric catalog
\footnote{http://alan-meert-website-aws.s3-website-us-east-1.amazonaws.com/fit\_catalog/index.html} 
\citep{Meert-15}, which contains $\sim 680,000$ galaxies from the SDSS DR7 spectroscopic sample. 
We adopt the measurement with \sersic\ bulge + exponential disk model. 
The model of \sersic\ bulge is more flexible than the model of de Vacouleurs bulge, 
and is likely to be a better choice for both high and low mass galaxies. 

Our final sample contains 524,852 galaxies, of which 24\% are satellite galaxies. 
Since this is a flux limited sample, we assign each galaxy a weight 
$w=(V_{\rm max}C)^{-1}$ to correct for the selection effect, where $V_{\rm max}$ is 
the co-moving volume between the minimum redshift and the maximum redshift to which the 
galaxy can be observed in the flux-limited survey \citep{Blanton-Roweis-07}, 
and $C$ is the spectroscopic completeness.

\subsection{L-GALAXIES and EAGLE}
\label{subsec:2.2}   

As already mentioned in Section \ref{sec:introduction}, in this
  paper we follow the methodology of \citetalias{Wang-18b}, and compare the
  observational results inferred from the SDSS data with predictions
  from the latest version of the Munich semi-analytical model,
L-GALAXIES\footnote{http://galformod.mpa-garching.mpg.de/public/LGalaxies/}
\citep{Henriques-15, Henriques-17} and the state-of-the-art
hydrodynamic simulation,
EAGLE\footnote{http://eagle.strw.leidenuniv.nl/} \citep{Crain-15,
  Furlong-15, Schaye-15, McAlpine-16}. This allows us to fairly and
directly interpret the observational results in the context of galaxy
formation models. In particular, it tests to what extent the
environmental dependence of the size and structural evolution of
galaxies in EAGLE and L-GALAXIES is compatible with observations.

Here we briefly describe the two models; the details of the two models
and how they are used to construct the mock catalogs used here can
be found in \citetalias{Wang-18b}. Semi-analytic models are phenomenological models
that take the advantage of empirically motivated prescriptions to
describe baryonic processes, such as gas accretion, cooling and
heating, star formation, stellar and AGN feedback, and
tidal/ram-pressure stripping.  As the latest version of the Munich
model, L-GALAXIES is built upon the Millennium Simulation
\citep{Springel-05} and employs a Markov Chain Monte Carlo method to
explore the high-dimensional parameter space to match the observed
galaxy stellar mass function and quenched fraction as a function of
stellar mass from redshift of 0 to 3. In L-GALAXIES, the disk size of
a galaxy is determined by the angular momentum of its infalling cold
gas,
while three different mechanisms are invoked to grow a bulge:
major mergers, minor mergers and disc instability. In
  particular, during a major merger the disks of the progenitors are
assumed to be completely destroyed and a spheroidal galaxy
forms. During a minor merger, the stars from the least
  massive progenitor are added to the bulge component, while the
  stellar disk of the more massive progenitor remains unchanged.
Note that, in L-GALAXIES, mergers among satellite galaxies are
  rare; the majority of all mergers are between a central and one of
  its satellites \citep{Guo-11, Henriques-15}. Tidal and ram-pressure
stripping are assumed to affect only the hot gas surrounding satellite
galaxies. Hence, these `satellite-specific' processes do not
  {\it directly} affect the stellar distributions of satellite
  galaxies, but they do suppress their subsequent growth, and thus
  their size and structural evolution, by quenching star
  formation. Because of this specific treatment of processes that only
  operate on galaxies identified by the code as satellites, the
  structural properties of centrals and satellites in L-GALAXIES are
  expected to be shaped by different processes.

The EAGLE simulation adopts advanced smoothed particle hydrodynamics
and subgrid models for a series of baryonic processes, such as gas
cooling, metal enrichment, black hole growth, and stellar and AGN
feedback. Free parameters in the feedback models are tuned by matching
the galaxy stellar mass function and the stellar mass-black hole mass
relation at $z\sim 0$ \citep{Crain-15, Furlong-15}.  The luminosities
and stellar masses of EAGLE galaxies are obtained from their star
formation histories by assuming the \cite{Chabrier-03} initial mass
function and the \cite{Bruzual-Charlot-03} stellar population model.
In contrast to L-GALAXIES, EAGLE adopts the same subgrid prescriptions
for both centrals and satellite, and so the differences between these
two populations, if any, must be due to their different environments.  
For instance, environmental effects such as tidal or ram-pressure stripping are 
treated self-consistently by the gravity and hydrodynamics solvers, 
without explicitly taking into consideration whether the galaxy is a central or a satellite.
Following \citetalias{Wang-18b}, we use the Ref-L100N1504 simulation, which has a
box size of 100 Mpc sampled with $2 \times 1504^3$ particles.  The
simulation contains more than 11,000 dark matter halos with masses
above $10^{11}$\msun, and nearly 10,000 galaxies with masses
comparable to or larger than that of the Milky Way.

In both L-GALAXIES and EAGLE, halo mass is defined as the mass within the radius
corresponding to an overdensity of 200. This halo mass is only slightly different
from that defined in \cite{Yang-07} for halos with a NFW profile 
\citep[Navarro-Frenk-White;][]{Navarro-96}. The stellar masses (recommended) in EAGLE 
are measured within a typical aperture of 30 kpc, to avoid the contamination of 
inter-cluster/intra-group light. The mass loss due to this effect is negligible for 
low-mass galaxies. However, for more massive galaxies the aperture reduces the stellar 
masses somewhat by cutting out intra-cluster light. At a stellar mass of $10^{11}$\msolar, 
the mass loss due to aperture effect is only 0.1 dex. This is only a minor effect for 
the stellar mass range we are considering, and similar for both centrals and satellites.

The structural parameters for the model galaxies are also taken from the publicly 
released data. Unfortunately, the publicly available galaxy catalogs 
L-GALAXIES and EAGLE do not contain both size and bulge-to-total ratio. 
EAGLE only provides the half-mass radii, while L-GALAXIES only bulge-to-total mass ratios.
We should also keep in mind that the parameters for model galaxies are measured in a 
very different way from those in the observational data. 
In principle, one may construct the $r$-band images of model galaxies to fully mimic 
the observation, and measure the structural parameters of model galaxies using the same 
methods as in the observation. However, this is clearly beyond the scope of 
the present paper. Thus, we can only present a qualitative comparison between 
the model predictions and observational data.

To facilitate the model-data comparison, we construct mock galaxy catalogs for both L-GALAXIES and EAGLE that properly account for various observational effects \citep[for details, see][and \citetalias{Wang-18b}]{Lim-17}. We use the snapshots at $z=0.1$ (the median redshift of SDSS galaxies used in this paper) to construct our mock catalogs. L-GALAXIES uses a simulation box of $480.3 h^{-1}{\rm Mpc}$ on a side, while the box size of the EAGLE simulation is only 100 Mpc. Both are significantly smaller than the volume probed by the SDSS, and we therefore stack duplicates of the original simulation boxes side by side to construct a sufficiently large volume.  We then choose a location for the observer, and calculate the redshift and the apparent magnitude for each model galaxy based on its luminosity, distance, and velocity with respect to the observer. Finally, we select a flux-limited sample of galaxies from a light cone covering the redshift range $0.01<z<0.2$, which is similar to that covered by our SDSS group catalog. All comparisons are based on these catalogs unless specified otherwise. Finally, we note both L-GALAXIES and EAGLE adopt the $Planck$ cosmology \citep[][]{Planck-14a, Planck-14b}, which differs somewhat from the WMAP3 cosmology \citep{Spergel-07} adopted in the construction of our SDSS group catalog. We note that the different cosmologies is not a concern as long as the results are presented in a self-consistent way, i.e. we use the WMAP3 cosmology for the analysis of the SDSS group catalog, while use $Planck$ cosmology for the analysis of the EAGLE and L-GALAXIES group catalogs. 


\section{Results}
\label{sec:results}

Massive galaxies are typically larger in size and more bulge-dominated
than less massive ones \citep[e.g.][]{Simard-11, vanderWel-14}. Hence,
it is important to control the stellar mass when comparing centrals
to satellites. We do this by focusing on the  \mstar-\re\ and
\mstar-B/T relations.
Specifically, we investigate whether central and satellite galaxies
follow the same relations and how these relations depend on
environmental parameters, such as halo mass and halo-centric distance.
\begin{figure*}
  \begin{center}
    \epsfig{figure=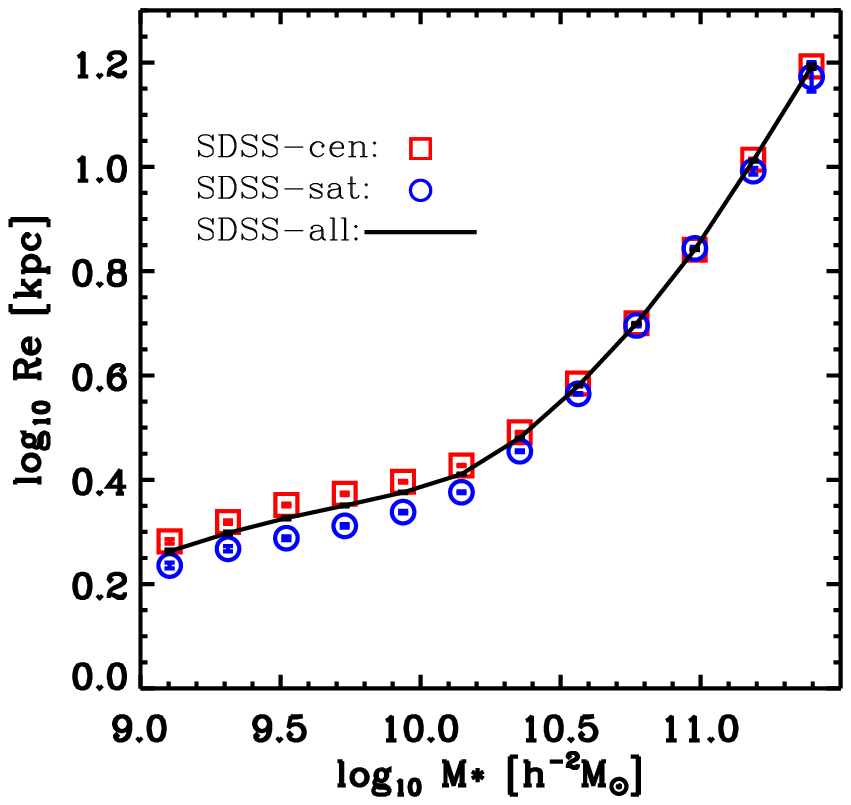,clip=true,width=0.40\textwidth} 
    \epsfig{figure=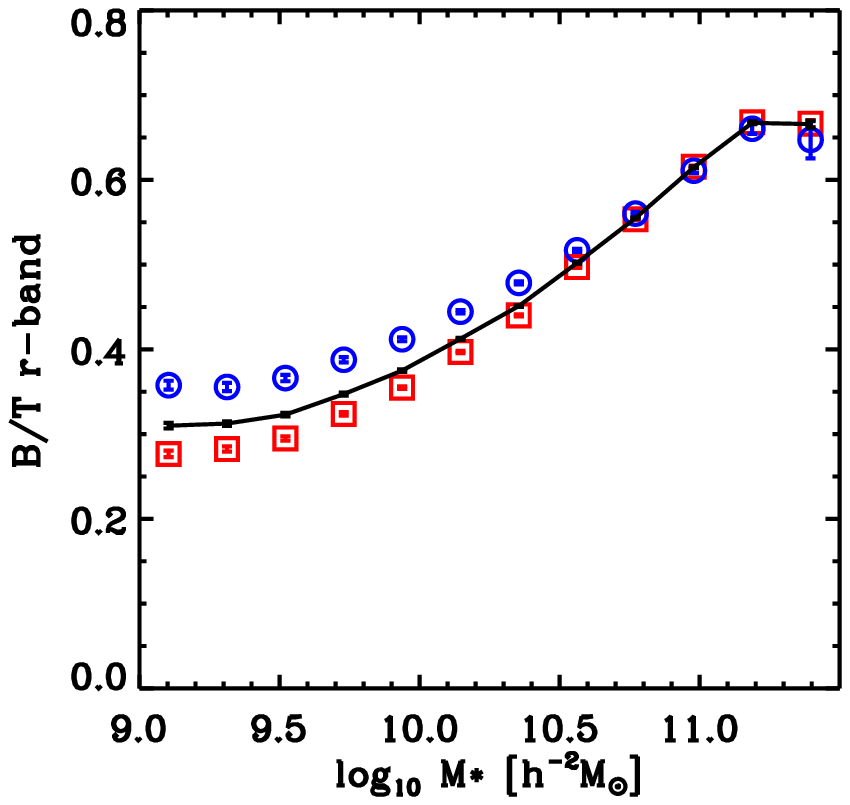,clip=true,width=0.40\textwidth} 
    \end{center}
  \caption{\mstar-\re\ (left-hand panel) and \mstar-B/T (right-hand
    panel) relations for central, satellite and all galaxies, selected
    from SDSS galaxy and group catalogs.  In each panel, the results
    for centrals, satellites and all galaxies are indicated by red
    squares, blue circles and black line, respectively. }
  \label{fig:all_mstar_0th}
\end{figure*}

\begin{figure*}
  \begin{center}
      \epsfig{figure=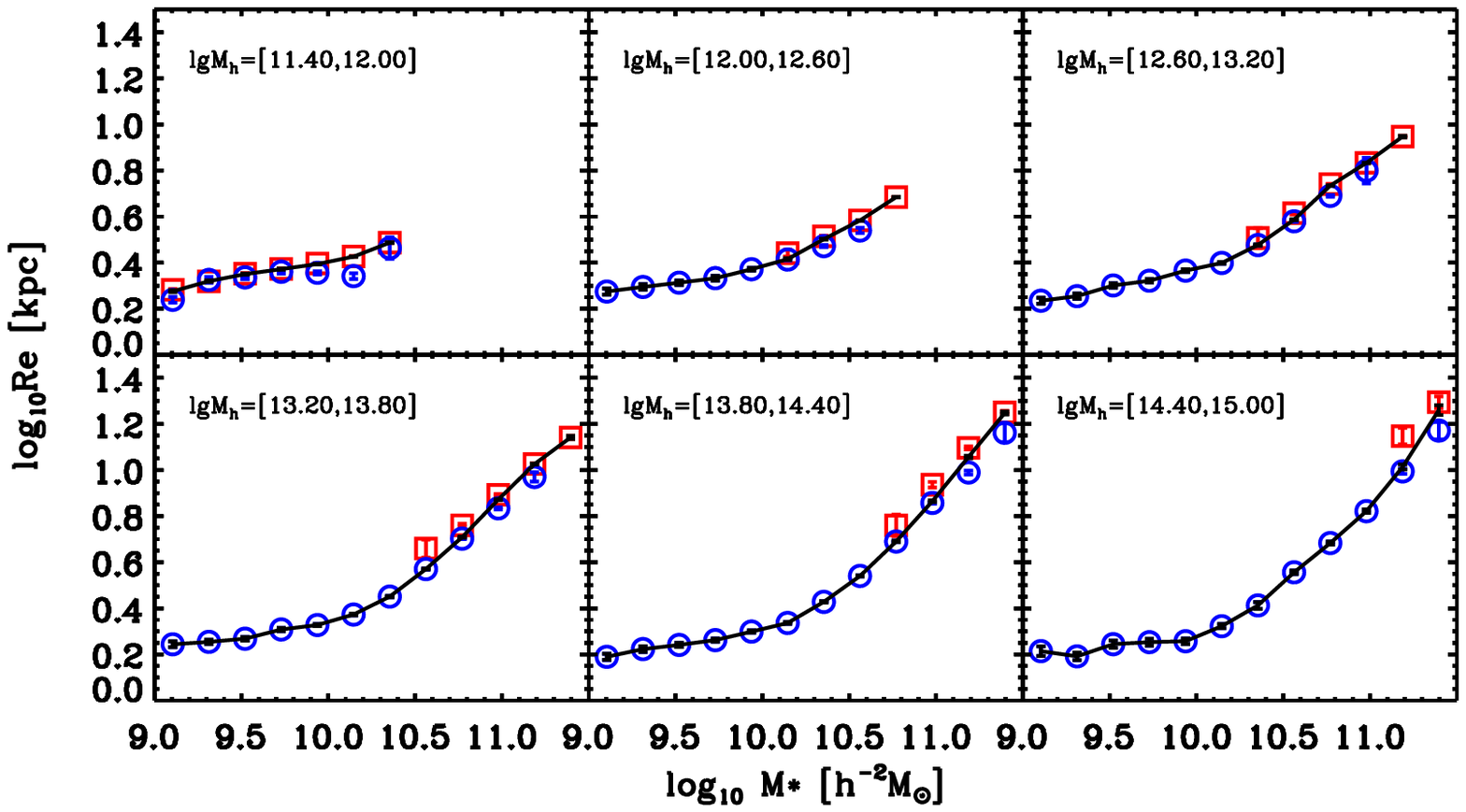,clip=true,width=0.8\textwidth}
    \epsfig{figure=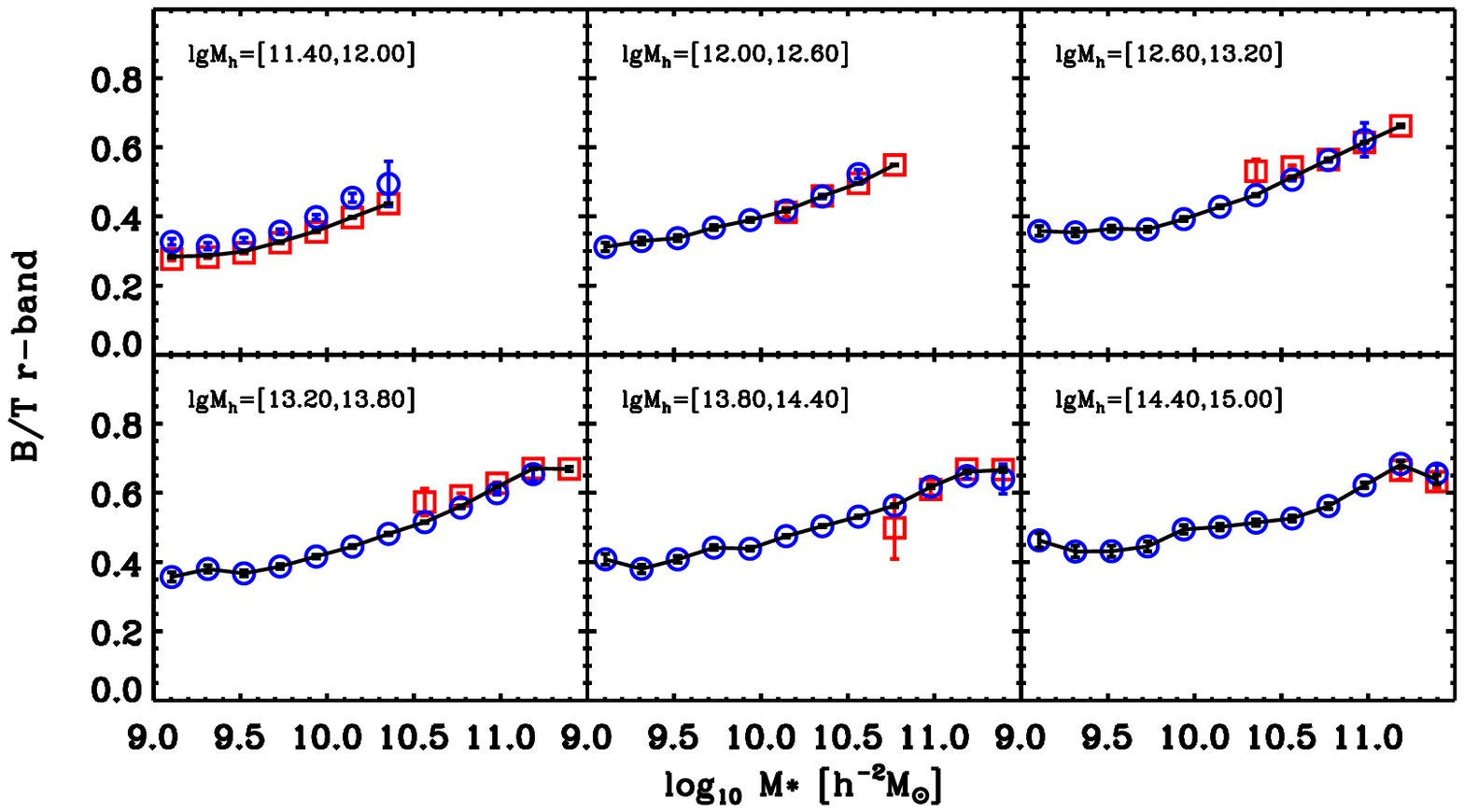,clip=true,width=0.8\textwidth} 
  \end{center}
  \caption{Top group of panels show \mstar-\re\ relations and bottom
    group of panels show \mstar-B/T relations for central and
    satellite galaxies in various halo mass bins.  As in Paper
    I, we separate galaxies into six halo mass bins, which have the
    same width of 0.6 dex in logarithmic space from
    $10^{11.4}h^{-1}$\msun\ to $10^{15.0}h^{-1}$\msun.  In each panel,
    the results for centrals, satellites and all galaxies are
    indicated by red squares, blue circles and black line,
    respectively. }
  \label{fig:struct_mass_1th}
\end{figure*}

\begin{figure*}
    \begin{center}
    \epsfig{figure=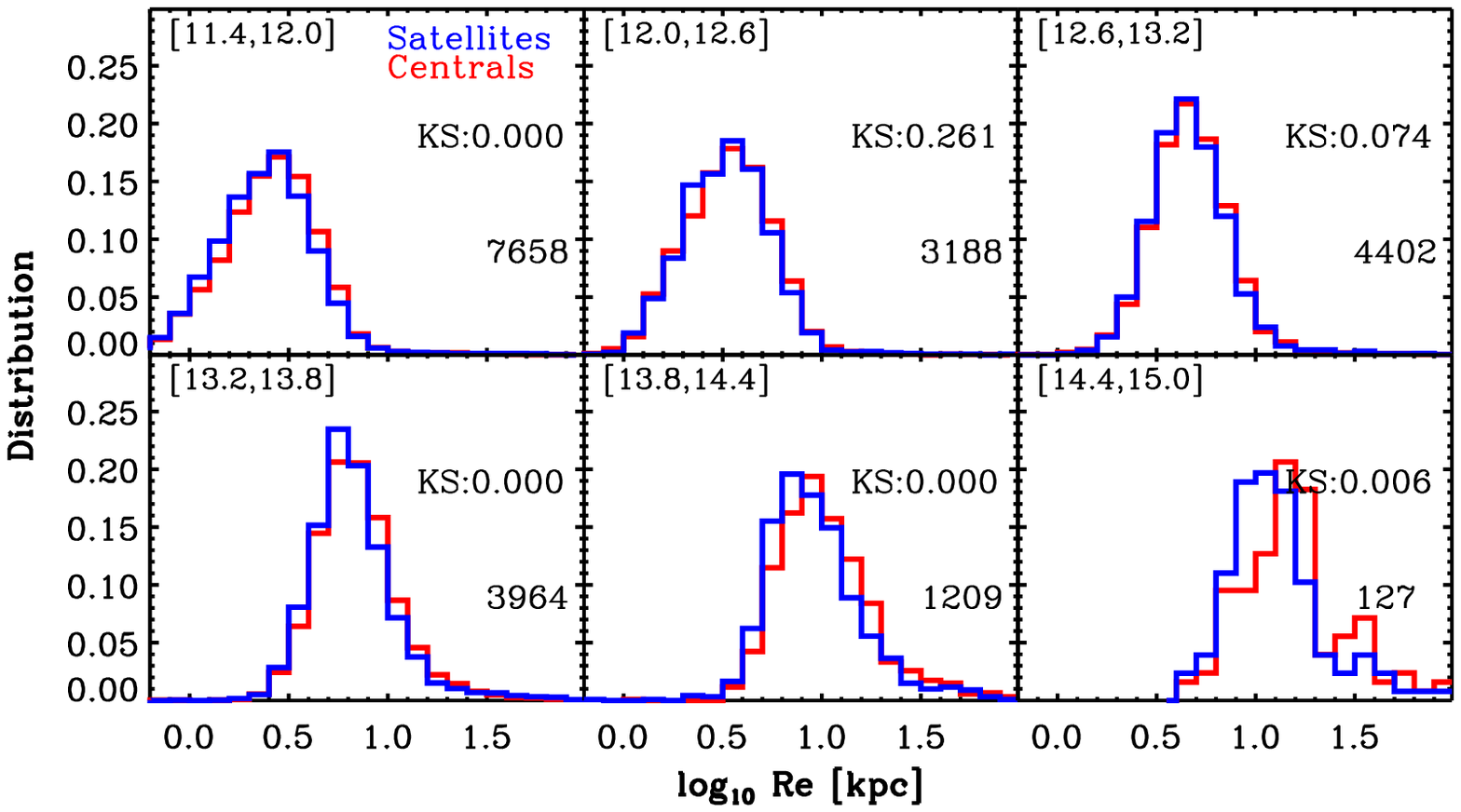,clip=true,width=0.80\textwidth} 
    \epsfig{figure=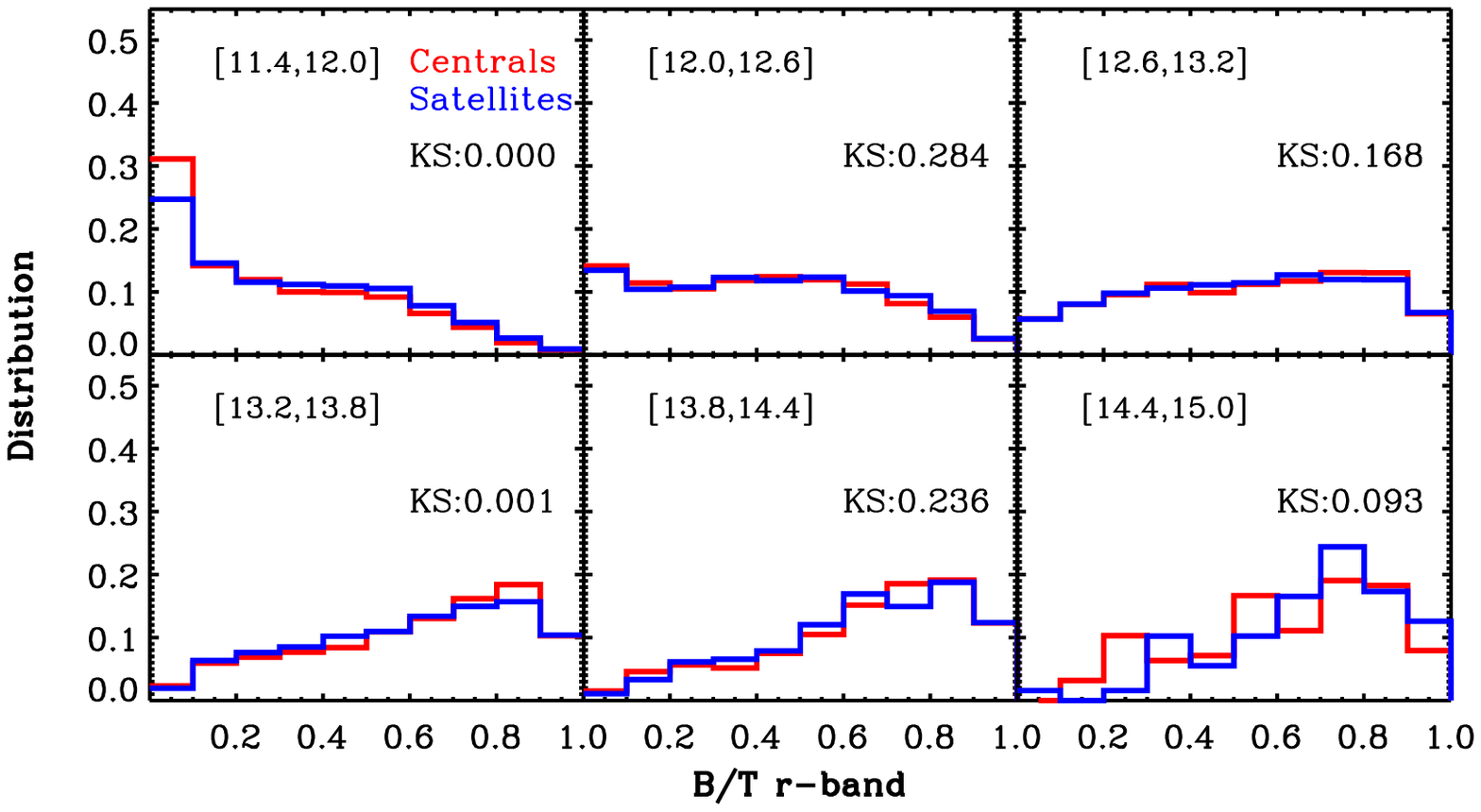,clip=true,width=0.80\textwidth} 
  \end{center}
\caption{ Top group of panels:  The distribution of size for centrals and satellites 
with controlled stellar mass at the six halo mass bins.  Bottom group of panel: 
The same as the top group of panels, but for the distribution of B/T. 
In top group of panels, we indicates the number of 
the centrals (or satellites) after controlling the stellar mass 
in the bottom right corner of each panel. After controlling the distribution
of stellar mass, the centrals and satellites have the same number of galaxies, 
at each halo mass bin. In each panel, we also denote the Kolmogorov-Smirnov 
test probability between the two distributions. 
}
\label{fig:distribution}
\end{figure*}

\begin{figure*}
    \begin{center}
    \epsfig{figure=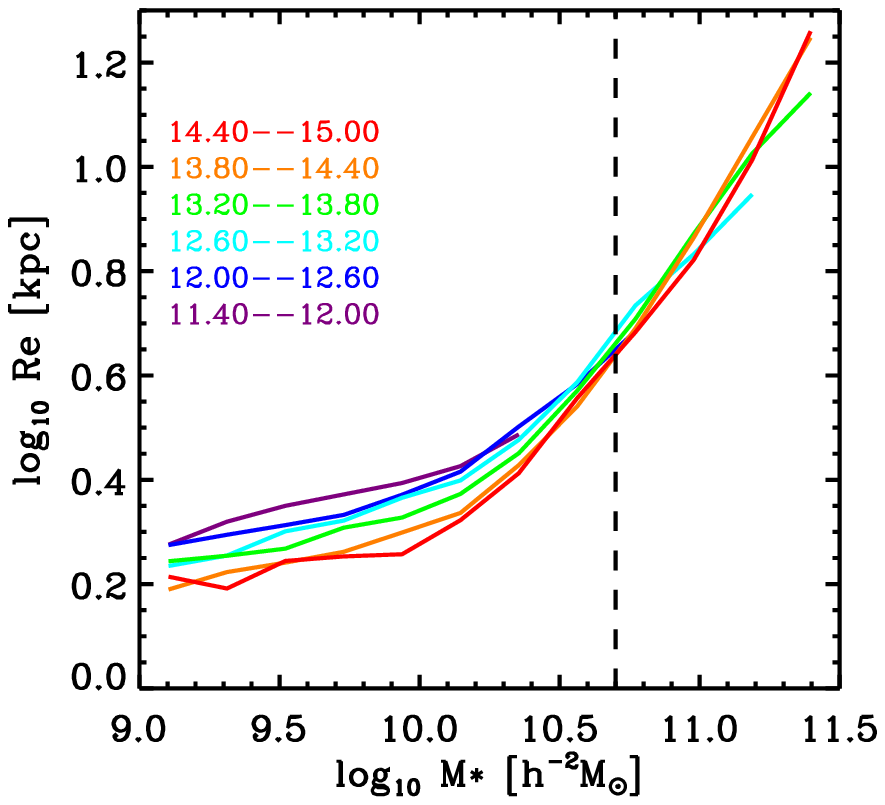,clip=true,width=0.40\textwidth} 
    \epsfig{figure=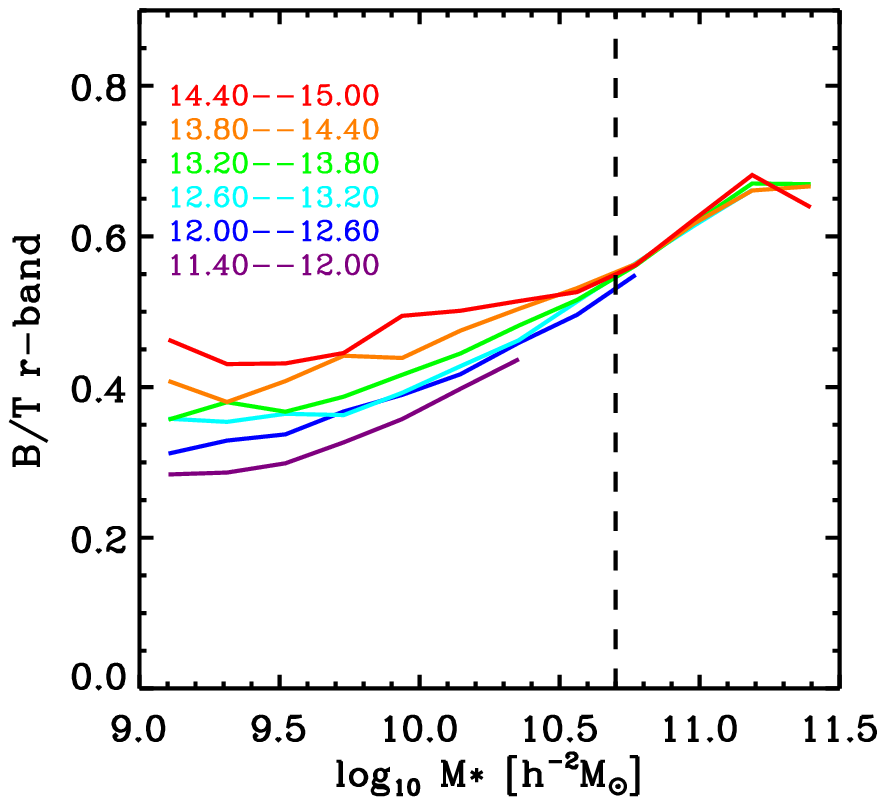,clip=true,width=0.40\textwidth} 
  \end{center}
\caption{\mstar-\re(left-hand panel) and \mstar-B/T relations
  (right-hand panel) for galaxies of different halo masses, as indicated
  in the top-left corner. }
\label{fig:ms_all_0th}
\end{figure*}

\subsection{The sizes of centrals and satellites}
\label{subsec:3.1}

The left panel of Figure \ref{fig:all_mstar_0th} shows the average
\re\ (in logarithmic space) as a function of stellar mass for centrals
(red squares), satellites (blue circles) and all galaxies (black line). 
Here, and throughout the remainder of this paper, the errors are estimated
using 1000 bootstrap samples. Note that the \mstar-\re\ relations
for centrals and satellites are quite similar at \lgmstar$>10.5$,
while significant differences are evident at lower mass, with
centrals being systematically larger, by $\sim 0.06\,{\rm dex}$,
than satellites of the same stellar mass.

These results may reflect the fact that centrals and satellites have
experienced different environmental processes. After all,
  satellites usually reside in more massive halos than centrals of the
  same stellar mass, and halo mass is known to be one of the primary
  environmental parameters regulating galaxy formation (in particular,
  galaxy quenching). To test the role of halo mass in controlling the
  sizes of galaxies of given stellar mass, we separate
  galaxies into six logarithmic bins in halo mass, each with a 
  width of 0.6 dex,
  and covering the range from $10^{11.4}h^{-1}$\msun\ to
  $10^{15}h^{-1}$\msun. For each bin, we again compute the
  \mstar-\re\ relations for centrals and satellites separately, the
  results of which are plotted in the top panels in Figure
\ref{fig:struct_mass_1th}. As one can see, the differences between the
two populations shown above are significantly reduced or even
eliminated: centrals and satellites follow basically the same
\mstar-\re\ relation at a given halo mass. This indicates that the
environmental processes that affect the sizes of centrals and
satellites are similar, strengthening the proposition that the
difference in the \mstar-\re\ relation for the two populations shown in
the left panel of Figure \ref{fig:all_mstar_0th} is primarily due to differences 
in the distributions of halo mass for centrals and satellites. 


Figure \ref{fig:distribution} shows the size 
distributions of centrals and satellites in different halo 
mass bins. To make a fair comparison, the two populations are 
controlled to have the same distribution in stellar mass. 
As one can see, the size distributions of centrals and satellites
are quite similar. For the two intermediate halo mass bins 
($10^{12.0}M_{\odot}h^{-1}<M_{\rm h}<10^{13.2} M_{\odot}h^{-1}$), the 
two distributions are almost indistinguishable, as indicated by  
the probabilities of the Kolmogorov-Smirnov test. For the other
three massive halo mass bins, there are small differences between 
the two populations, which may due to the fact that the halo mass
distribution of the two populations are not exactly the same.
The results for the most massive bin are noisy, 
as there are only about 100 galaxies in each population. 

Because galaxy size does not seem to depend on whether a galaxy is
classified as a central or a satellite, we calculate the average
\mstar-\re\ relations for galaxies in the six halo mass bins without
distinguishing centrals and satellites. The result is shown in the
left panel of Figure \ref{fig:ms_all_0th}.  As can be seen, galaxy
size decreases with increasing halo mass at the low stellar mass
end. This trend becomes less obvious with increasing stellar mass, and
almost disappears at \mstar$>10^{10.7}h^{-2}$\msun, as indicated by
the vertical dashed line.  The fact that galaxies of low stellar mass
are smaller in more massive halos may be due to a number
of processes. 
For example, more massive halos may be more effective in quenching star formation,
particularly for low-mass galaxies \citep[e.g.][]{Peng-10, Woo-15,
  Wang-18c}, probably due to stronger ram-pressure stripping and/or
shock-excited heating.  Consequently, low-mass galaxies in massive
halos may stop growing their sizes due to quenching, and the fading of the 
disk may cause the half-light radius to shrink
\citep{Lilly-Carollo-16}. Tidal stripping and galaxy harassment can
strip stars from the outer stellar disk of low-mass galaxies, making
the galaxy more compact and smaller. These two processes  
primarily depend on the mass density of the host halo,
which is similar for halos of different mass.


\subsection{The bulge-to-total light ratios of centrals and satellites}
\label{subsec:3.2}

In this subsection, we focus on the bulge-to-total light ratio (B/T)
to examine how environment impacts bulge formation in centrals
  and satellites.  As in Section \ref{subsec:3.1}, we first show the
mean B/T as a function of stellar mass in the right-hand panel of
Figure \ref{fig:all_mstar_0th}, without separating galaxies into
different halo mass bins. As one can see, at the low-mass end
satellites tend to harbor more pronounced bulges than
centrals. As with the sizes, the differences become
weaker with increasing stellar mass and vanish for
  \lgmstar$>10.5$.

To reduce effects caused by the difference in host halo masses between
centrals and satellites, we show B/T as a function of stellar mass for
centrals and satellites separated in six halo mass bins in the bottom
panels of Figure \ref{fig:struct_mass_1th}.  We see again that the
significant difference between the two populations shown in Figure
\ref{fig:all_mstar_0th} disappears when the halo mass is controlled.
This indicates that both centrals and satellites have similar B/T when
both the stellar mass and halo mass are controlled, indicating again
that centrals are not special in comparison with satellites as far as
their star formation (\citetalias{Wang-18a} and \citetalias{Wang-18b}) 
and structural properties are concerned.

Since there is no significant evidence to suggest that centrals and
satellites have systematically different B/T, we calculate the average
$M_*$-B/T relation for galaxies in the six halo mass bins without
distinguishing centrals from satellites. This is shown in the
right-hand panel of Figure \ref{fig:ms_all_0th}.  Clearly, galaxies
appear to have higher B/T in more massive halos for stellar masses
less than $10^{10.7}h^{-2}$\msun, but the environmental dependence
becomes very weak for galaxies above this stellar mass. This is
broadly consistent with previous studies \citep[e.g.,][]{Bamford-09,
Liu-19, Bluck-19}.


The B/T distributions of centrals and satellites, 
controlled to have the same stellar mass distribution,
are shown in the bottom group panels of Figure \ref{fig:distribution}. 
The sub-samples here are exactly the same as those shown in the upper 
panels. As one can see, centrals and satellites have a similar distribution 
in B/T for almost all the halo mass bins. 
Except for the two left panels ($10^{11.4}M_{\odot}h^{-1}<M_{\rm h}<10^{12.0} M_{\odot}h^{-1}$ and 
$10^{13.2}M_{\odot}h^{-1}<M_{\rm h}<10^{13.8} M_{\odot}h^{-1}$ ), 
the Kolmogorov-Smirnov test, in each of other panels, shows that the 
difference between the two populations is statistically insignificant.
This strengthens the conclusion 
that centrals and satellites have similar B/T when both stellar mass 
and halo mass are controlled. 

Combined with the result of Section \ref{subsec:3.1}, we
conclude that centrals and satellites have similar sizes and B/T when
both stellar mass and halo mass are controlled.  We have also examined
other structural properties than B/T, including the
\sersic\ index, the concentration, defined as the ratio of the radii
enclosing 90 and 50 per cent of the galaxy light, and the stellar
surface densities inside \re. All these show the same trends as B/T.  
As an example, we present the results for the stellar 
surface density within \re\ in the Appendix.  
Because of this, we therefore only present results based on 
B/T in the following.

A potential concern is that the results may be 
contaminated by groups that are not yet relaxed
and within which satellites may not have
experienced the environmental effects of their current host halos
\citep[e.g.][]{Carollo-13b}.  A significant fraction of groups
($\sim$6.0\% of groups, corresponding to $\sim$15.5\% of galaxies)
is found to have centrals that have projected halo-centric distances 
greater than $0.1 r_{\rm 180}$.  These groups are more likely to be 
un-relaxed and including them may reduce the difference between centrals 
and satellites. We have reexamined our results by excluding these groups,
and found no significant changes in our conclusion.

\subsection{Dependence on halo-centric radius}
\label{subsec:3.3}
\begin{figure*}
   \begin{center}
    \epsfig{figure=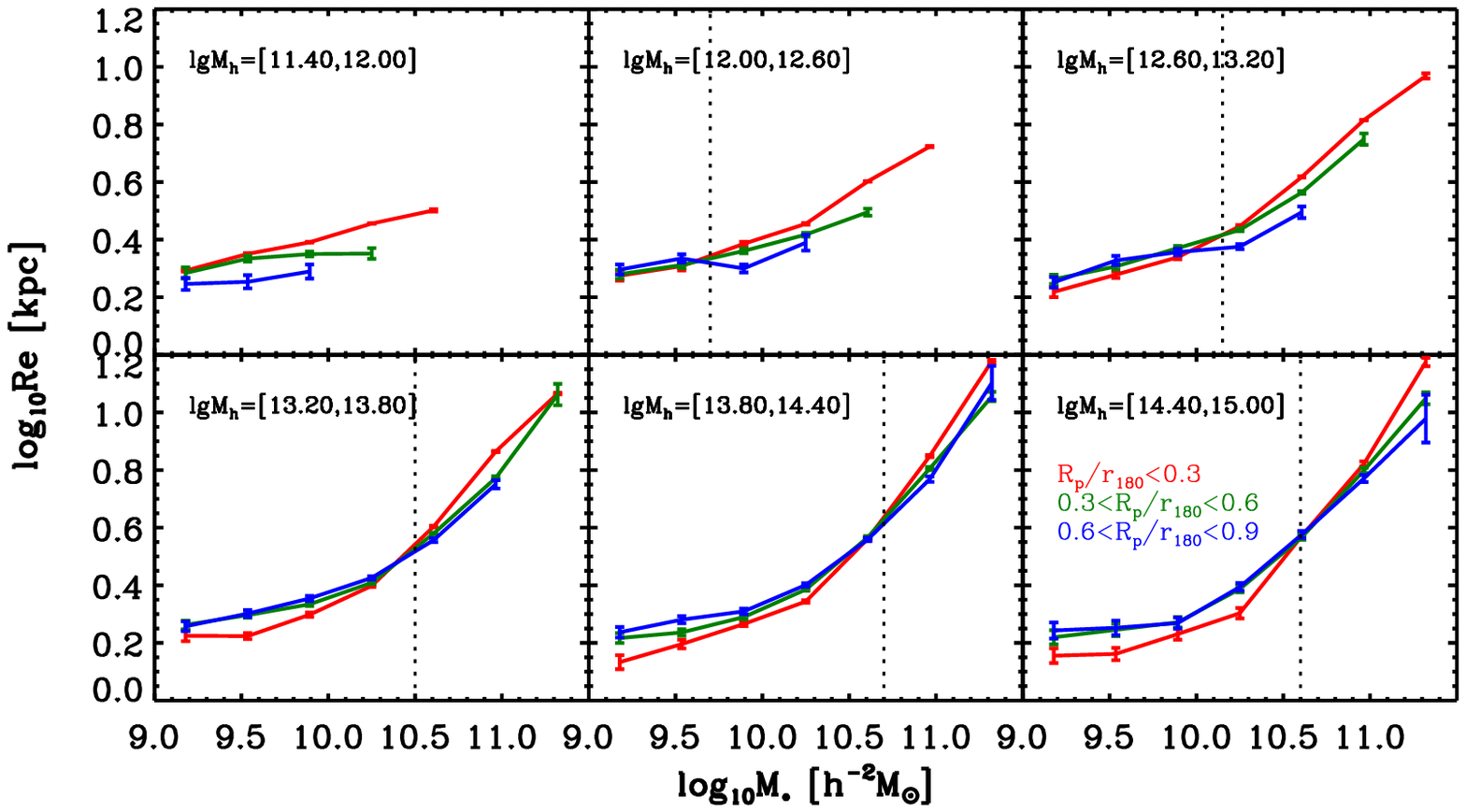,clip=true,width=0.8\textwidth} 
    \epsfig{figure=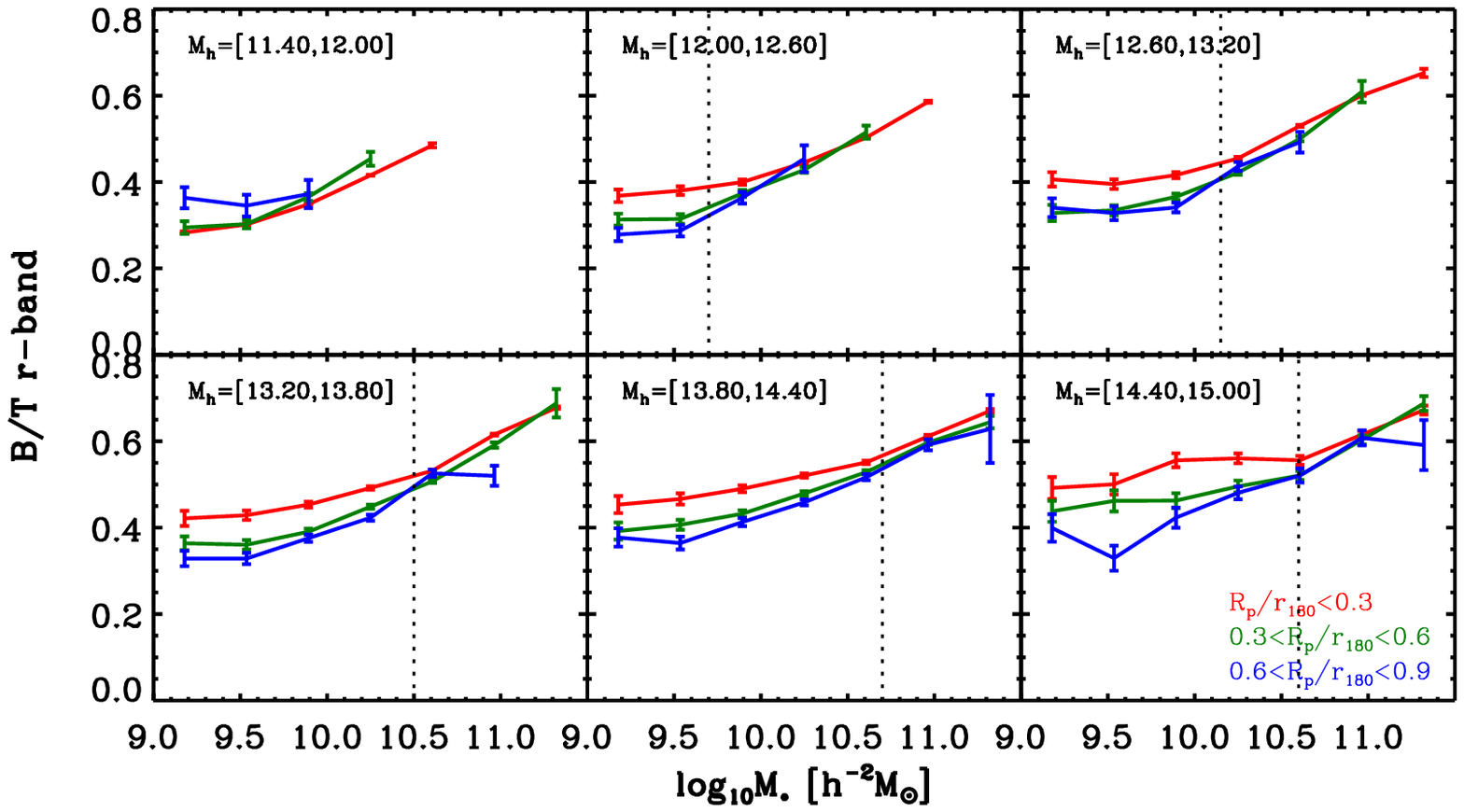,clip=true,width=0.8\textwidth} 
   \end{center}
  \caption{Top group of panels: \mstar-\re\ relations for galaxies
    with different halo mass (as indicated in each panel) and
    different halo-centric radii (as indicated in the bottom right
    panel).  Bottom group of panels: \mstar-B/T relations for galaxies
    with different halo mass and different halo-centric radii.  Note
    that we do not distinguish between centrals and satellites, since
    centrals and satellites show similar \mstar-\re\ and \mstar-B/T
    relations as long as one controls halo mass (see Section
    \ref{subsec:3.1} and \ref{subsec:3.2}). The vertical dashed lines
    show the transitional stellar mass for star formation quenching,
    taken from \citetalias{Wang-18a}.
}
  \label{fig:pos_mh_1th}
\end{figure*}

\begin{figure}
\begin{center}
\epsfig{figure=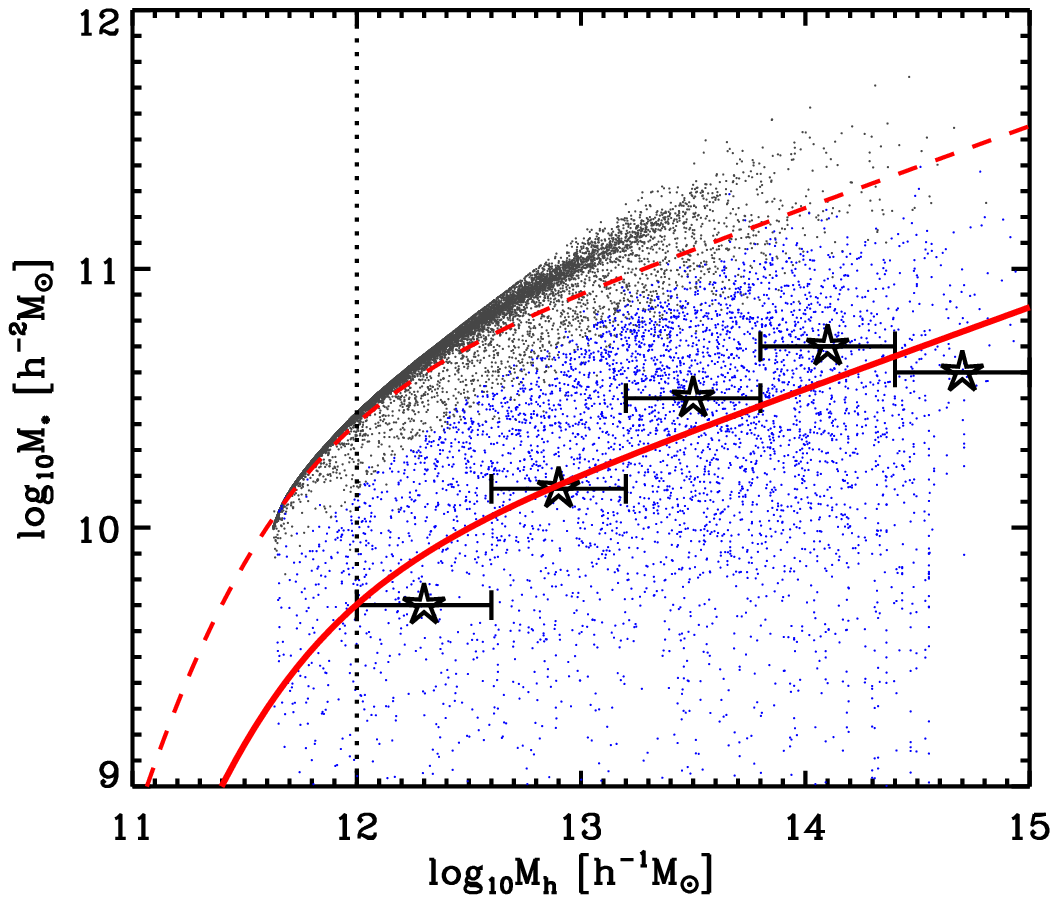,clip=true,width=0.48\textwidth} 
\end{center}
\caption{Stellar mass, $M_*$, as function of halo mass, $M_{\rm
      h}$.  Pentagrams indicate the transitional stellar masses for 5
    different halo mass bins, with error bars indicating the bin
    widths. Small dots are 20,000 galaxies randomly sampled from the
    SDSS group catalog, with centrals in black and satellites in
    blue. The dashed, red curve is the $M_*-M_{\rm h}$ relation for
  central galaxies taken from \cite{Yang-Mo-vandenBosch-09},
  while the solid line is the same relation by offset by 0.7
    dex, which nicely matches the transitional stellar mass as
    function of halo mass.  The vertical dotted line shows the
  $M_{\rm h}=10^{12}h^{-1}{\rm M_{\odot}}$, below which we do not
  see a transitional stellar mass. 
  }
  \label{fig:theshold}
\end{figure}

Various environmental processes discussed in the literature, such as
tidal stripping, ram pressure stripping, strangulation, and
harassment, are all expected to depend on the location of a galaxy
within its host halo.  Indeed, recent analyses have revealed that the
quenched population of satellite galaxies becomes more dominant toward
the halo center \citep[e.g.][]{Weinmann-06, vandenBosch-08,
  Wetzel-Tinker-Conroy-12, Kauffmann-13, Woo-13}. This either
  indicates that the processes associated with quenching
become stronger towards the halo center, and/or it
  reflects the fact that satellites that reside closer towards the
  center where accreted earlier \citep[e.g.,][]{Gao-04, Contini-12,
    vdBosch-16}, and have therefore been exposed to quenching-inducing
  processes for a longer duration. Here we examine how the structural
properties depend on the halo-centric distance ($R_{\rm p}$). We do
not discriminate centrals and satellites, since the two populations
intrinsically have no differences in size and structural properties
(see Section \ref{subsec:3.1} and \ref{subsec:3.2}).

Figure \ref{fig:pos_mh_1th} shows the \mstar-\re\ and \mstar-B/T relations for galaxies in three halo-centric distance bins: $R_{\rm p}/r_{\rm 180}<0.3$ (red lines), $0.3<R_{\rm p}/r_{\rm 180}<0.6$ (green lines), and $0.6<R_{\rm p}/r_{\rm 180}<0.9$ (blue lines), where $r_{\rm 180}$ is the halo virial radius \citep[see][]{Yang-07}. Both the size and B/T depend on $R_{\rm p}/r_{\rm 180}$ over the whole halo mass range. 

There are two interesting phenomena. The first one is that the $R_{\rm p}/r_{\rm 180}$ dependences seem to be driven mainly by the innermost bin. This suggests an existence of a critical halo-centric radius, beyond which the mass-size (or mass-B/T) relation does not depend on the halo-centric radius. We search for the critical halo-centric radius (if any) by dividing galaxies into more halo-centric bins. We find that the mass-size (and the mass-B/T) relation does not depend on the halo-centric radius at $R_{\rm p}/r_{\rm 180}$ above $\sim$0.5, at least for the three highest halo mass bins. It is not surprising for the existence of critical halo-centric radius, since beyond this radius most of galaxies may have not fallen into the inner region of the halos and therefore have not yet been affected by the environmental effects.

The second one is that the dependence of the size on $R_{\rm p}/r_{\rm 180}$ is different for low and high mass galaxies. 
Except for the lowest halo mass bin, low-mass galaxies appear to have larger size in the outer 
region of halos than those of the same mass in the inner region. 
This trend is reversed at the high \mstar\ end.  Similar results can be 
seen in B/T: at low \mstar\, galaxies at smaller projected halo-centric distances 
tend to have larger B/T. At the high
  \mstar\ end, though, this dependence of B/T on $R_{\rm p}$ vanishes.
This motivates us to define a `transitional' stellar mass,
$M_{\rm *,t}$, above which the dependence of the size and B/T 
on $R_{\rm p}$ is absent or becomes opposite to that at the 
low stellar mass end.  As is evident from Figure \ref{fig:pos_mh_1th}, 
$M_{\rm *,t}$ increases with halo mass, consistent with the fact that
some of the environmental processes are expected to be 
stronger in more massive halos. 
Furthermore, comparing the upper and lower sets of panels
shows that the size and B/T have similar transitional masses for given
halo mass, indicating that the changes in size and B/T
may be driven by similar processes.

In \citetalias{Wang-18a}, we found that the quenched fraction of low-mass galaxies
strongly depends on $R_{\rm p}$, and the dependence becomes weak at
the high mass end \citep[see also][]{Wang-18c}.  The transitional
stellar masses for star formation quenching, taken from \citetalias{Wang-18a}, are
shown by the vertical dotted line in each panel of Figure
\ref{fig:pos_mh_1th}.  It is interesting to see that $M_{\rm *,t}$ for
star formation quenching is very similar to those for the size and
B/T.  This provides strong support that star formation quenching, size
and structural evolution of galaxies may be connected. We have also
examined the dependence of the inner stellar mass surface density
within \re\ ($\Sigma_{\rm Re}$) on the scaled
halo-centric radius at given stellar mass (see Appendix).
The behavior of $\Sigma_{\rm Re}$ is quite similar to that of B/T 
(Figure \ref{fig:pos_mh_1th}), with a similar transitional stellar mass 
for each halo mass bins. 


The pentagrams in Figure \ref{fig:theshold} show the
transitional stellar mass as a function of halo mass, with the error
bars indicating the halo mass bins. Grey and blue dots indicate
  centrals and satellites from a sample of 20,000 galaxies randomly
  selected from the SDSS group catalog, while the red, dashed curve
  indicates the central mass-halo mass relation obtained by
  \cite{Yang-Mo-vandenBosch-09}, which is given by
\begin{equation}
M_{\rm *,t} = M_{\rm 0}\frac{(M_{\rm h}/M_{\rm 1})^{\alpha+\beta}}{(1+M_{\rm h}/M_{\rm 1})^{\beta}}, 
\label{eq:trans}
\end{equation}
with $\log_{10} M_{\rm 0}=10.31$, $\log_{10}M_{\rm 1}=11.04$,
$\alpha=0.3146$ and $\beta=4.5427$.  For comparison, the red, solid
curve is the same relation but shifted downwards by 0.7 dex (i.e.,
with $M_0 \rightarrow 0.2 M_0$), which is a reasonably good fit to the
$M_{\rm *,t}-M_{\rm h}$ relation, indicating that $M_{\rm *,t}$ is
approximately one-fifth of the stellar mass of the average central
galaxy at the corresponding halo mass. Interestingly, a similar
  transitional mass was identified by \cite{Wang-18c}, who found that
the dependence of quenched fraction on halo-centric distance becomes
insignificant for galaxies with stellar masses greater than one fifth
of the masses of their centrals. 
From this figure, we can clearly see that the transition 
behavior is totally determined by satellite galaxies, as
central galaxies are all above and far away from the transition curve.

To conclude, our results indicate that environmental effects on
the size and structural properties of galaxies, as well as on
star formation quenching, depend on where the galaxies are located on
the $M_*-M_{\rm h}$ diagram: galaxies located above the $M_{\rm
    *,t}-M_{\rm h}$ relation (the solid, red line in Figure
  \ref{fig:theshold}), have sizes, quenched fractions and structural
  properties that are all independent of their location with their
  host halos. For galaxies below the $M_{\rm *,t}-M_{\rm h}$ relation,
  though, significant dependences on $R_{\rm p}$ emerge. We will
discuss the implications of these findings in Section
\ref{sec:summary}.

\subsection{Uncertainties in decomposition and seeing effect}

The parameter B/T is highly model dependent. 
In our analysis above, we have adopted the measurements based 
on the \sersic\ bulge + exponential disk model of
the $r$-band image. In their original paper, \cite{Meert-15} have 
performed the 2-dimensional modelling with other models, such as 
the de Vacouleurs bulge + exponential disk model, and found 
that the decomposition results can differ from that 
based on the \sersic\ bulge + exponential disk model.
In \citetalias{Wang-18a}, we adopted the B/T measurements based on the 
de Vacouleurs bulge + exponential disk model from \cite{Simard-11}. 
In order to quantify whether or not our results are sensitive 
to the different measurements, we have made tests using the 
B/T measurements of both \cite{Simard-11} and \cite{Meert-15}
based on the de Vacouleurs bulge + exponential disk model. 
Our tests clearly showed that our results are insensitive to the 
use of different decomposition measurements.

Another important issue is the seeing effect.
Even though the size and B/T are measured from 
the PSF (point spread function) de-convolved light profiles,  
seeing effect may still influence the accuracy of the measurements 
of both the size and B/T, especially for small galaxies.
To check this effect quantitatively, we have made the 
same plots shown in Sections \ref{subsec:3.1}, \ref{subsec:3.2} 
and \ref{subsec:3.3} using a sub-sample limited to $z<0.1$, where 
the half-width at half-maximum of the SDSS PSF (0.7 arcsec) corresponds 
to $1.3\,{\rm kpc}$. We found little change in the results.
This is expected, as the seeing effect affects centrals and 
satellites in a similar way, and for most of the halo mass 
bins, the comparisons between centrals and satellites  
are for relatively massive galaxies (\mstar$>10^{10.0}h^{-2}$\msun).

\section{Comparison with L-GALAXIES and EAGLE}
\label{sec:model}

In this section, we compare the observational results presented above with the mock samples generated from L-GALAXIES and EAGLE. Note that for EAGLE galaxies, only half-mass radii are available, and L-GALAXIES only provides bulge to total mass ratio. 
Because of this, we compare galaxy sizes predicted by EAGLE and the B/T 
predicted by L-GALAXIES with the observational results, respectively.  
As described in \S\ref{subsec:2.2}, 
the estimates of the two parameters for model galaxies are  
different from that for SDSS galaxies, and the comparison between 
model and observation can only be qualitative. 
As long as the differences between centrals and satellites 
are not affected by the differences in the estimators,   
we can still study whether EAGLE or L-GALAXIES can reproduce 
the observed similarity between centrals and satellites 
and the observed characteristic masses in the dependence 
of the galaxy structure on the environment.

In \S\ref{subsec:4.1}, we examine whether the two models can reproduce the
similarity between centrals and satellites. In \S\ref{subsec:4.2}, we
examine the dependence of galaxy size and B/T on halo-centric radius. In
both subsections, halo masses and central-satellite classifications 
are taken directly from the models. In \S\ref{subsec:4.3} we investigate 
uncertainties that are introduced by the group finder.

\subsection{Galaxy size and B/T in galaxy formation models}
\label{subsec:4.1}

\begin{figure*}
   \begin{center}
    \epsfig{figure=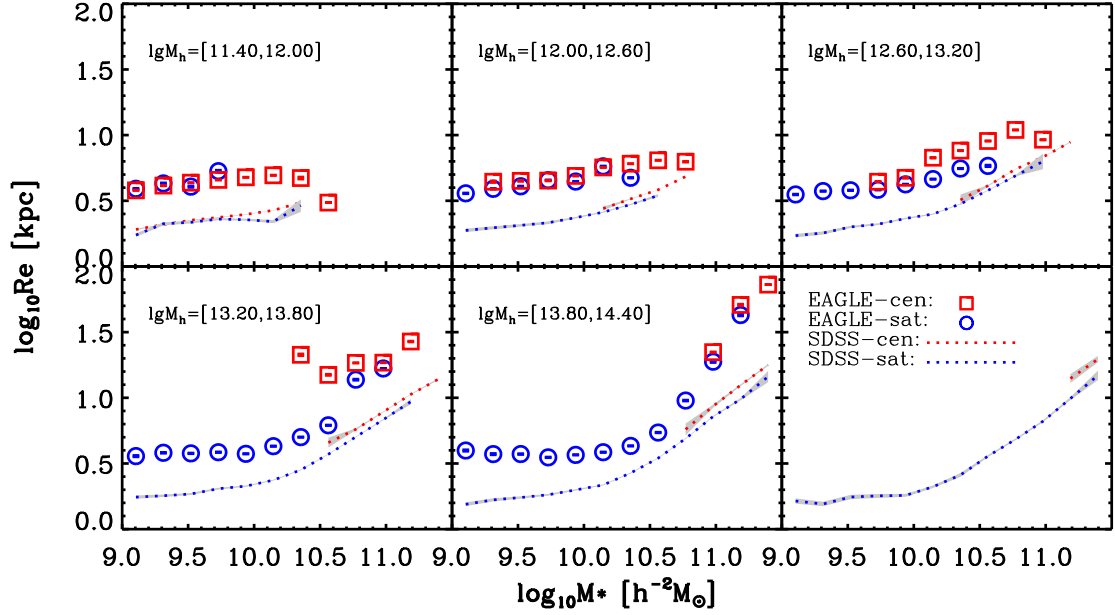,clip=true,width=0.90\textwidth} 
   \end{center}
  \caption{The half-mass radius as a function of stellar mass for
    centrals (red squares) and satellites (blue circles) in EAGLE. The
    results for SDSS centrals and satellites are also presented in
    dotted red and blue lines for comparison, taken from the top group
    of panels of Figure \ref{fig:struct_mass_1th}. 
    The grey, shaded regions indicate the 1$\sigma$ confidence range. }
  \label{fig:mstar_eagle_1th}
\end{figure*}

\begin{figure*}
   \begin{center}
    \epsfig{figure=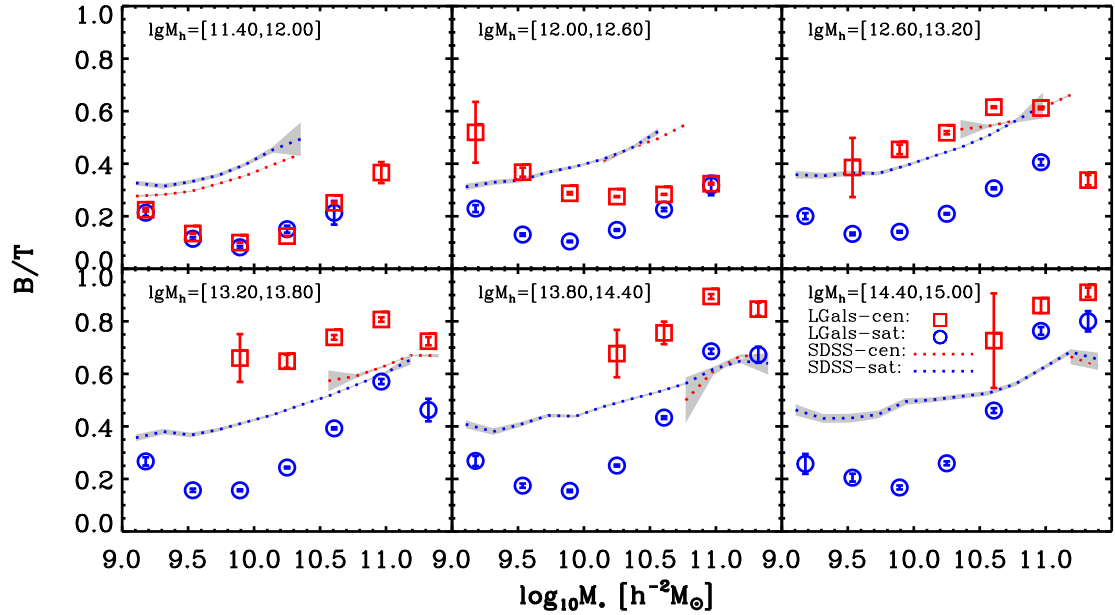,clip=true,width=0.90\textwidth} 
   \end{center}
  \caption{The bulge-to-total mass ratio as a function of stellar mass
    for centrals (red squares) and satellites (blue circles) in
    L-GALAXIES.  The \mstar-B/T relation for SDSS centrals and
    satellites are shown in dotted red and blue lines, taken from the
    bottom group of panels of Figure \ref{fig:struct_mass_1th}. The
    grey, shaded regions are the 1$\sigma$ confidence range. }
  \label{fig:mstar_lgals_1th}
\end{figure*}

Figure \ref{fig:mstar_eagle_1th} shows the half-mass radius of
centrals and satellites from EAGLE as a function of stellar mass in
six halo mass bins. Note that we use the same symbol, \re , to
  denote the half-mass radius as used previously to indicate the
  half-light radius of SDSS galaxies.  
Note that, because of the
relatively small volume of the EAGLE simulation, there are too few
massive halos to be able to plot any meaningful results for the most
massive halo mass bin ($14.4<\log_{\rm 10}M_{\rm h}/h^{-1}{\rm
M}_{\odot}<15.0$). We caution that this also implies that the
error bars for the massive EAGLE galaxies are severely underestimated
(see Section \ref{subsec:2.2} and \citetalias{Wang-18b}).

Overall, Figure \ref{fig:mstar_eagle_1th} reveals good,
  qualitative agreement with the SDSS results. In particular, for
  galaxies residing in halos with a mass less than
$10^{13.2}h^{-1}$M$_{\odot}$, the \mstar-\re\ relation of EAGLE
galaxies is nearly parallel to that of SDSS galaxies, albeit
with an overall offset of $\sim$0.3 dex.
The overall offset may be due to \textbf{different measurement methods} and the relatively bright surface brightness limit of the SDSS survey with respect to the EAGLE simulation, as demonstrated in \cite{Furlong-17}.
In massive halos, the offset becomes larger at 
both the low and high stellar mass ends. The larger offset 
at low mass ends may be due to the inadequate resolution, which blurs the boundaries of the low-mass galaxies, while the larger offset 
at high mass ends may be due to the inclusive of inter-cluster stellar components in the identification of EAGLE galaxies in groups or clusters.
  More importantly, centrals and satellites in EAGLE appear
to have similar \mstar-\re\ relations over the whole halo mass
range. An exception are the low-mass centrals in halos with
  $13.2 < \log_{\rm 10} M_{\rm h} / h^{-1}{\rm M}_{\odot} < 13.8$,
  which seem to have larger sizes than their corresponding
  satellites. However, there are only 7 unique (i.e., non-repeating)
  centrals in these two data points combined, indicating that the true
  statistical errors are much larger than those estimated from the
  mock data set, which uses many repetitions of the same simulation
  box. Hence, this apparent difference between centrals and satellites
  is not significant. 

We have used EAGLE to examine the size distributions 
  of centrals and satellites controlled in stellar mass for  
  a set of halo mass bins. We found that the two populations 
  have a similar distribution for halo masses less than 
  $10^{12.6}h^{-1}M_{\odot}$, while for higher halo masses
  the two size distributions have a small offset. This is consistent 
  with the result shown in Figure \ref{fig:mstar_eagle_1th}. 

Figure \ref{fig:mstar_lgals_1th} compares the B/T as a function
  of stellar mass for centrals (red squares) and satellites (blue
  circles) in L-GALAXIES to the SDSS results (dotted curves, taken
  from Figure \ref{fig:struct_mass_1th}). Clearly, the agreement is
  extremely poor; although L-GALAXIES roughly reproduces the
increasing trend of B/T with stellar mass, at least for
galaxies with \mstar\ $>10^{9.7}h^{-2}$\msolar, for less massive
  galaxies the trend predicted by L-GALAXIES is opposite to that in
  the data. In addition, typically there are large offsets between the
  average B/T predicted by L-GALAXIES and that observed for SDSS galaxies. 
  Most importantly, in halos with mass greater than
  $10^{12.0}h^{-1}$\msolar, centrals and satellites in L-GALAXIES
  typically have very different bulge-to-total ratios at fixed stellar
  and halo mass, inconsistent with the observational findings of
Section \ref{subsec:3.2}.

We emphasize that for SDSS galaxies, the size and B/T
  measurements are based on the light distribution, while for the
model galaxies they are based on the stellar mass
distributions. Although this could explain some of the offsets
  between the model and data, we are primarily concerned with the
  differences between centrals and satellites, and these differences
  should largely be unaffected by such offsets.
  We therefore conclude that, as a whole, EAGLE successfully
  reproduces the similar mass-size relations for centrals and
  satellites, while L-GALAXIES fails to reproduce the similar B/T of
  the two populations.

\subsection{The dependence on halo-centric radius}
\label{subsec:4.2}

\begin{figure*}
   \begin{center}
    \epsfig{figure=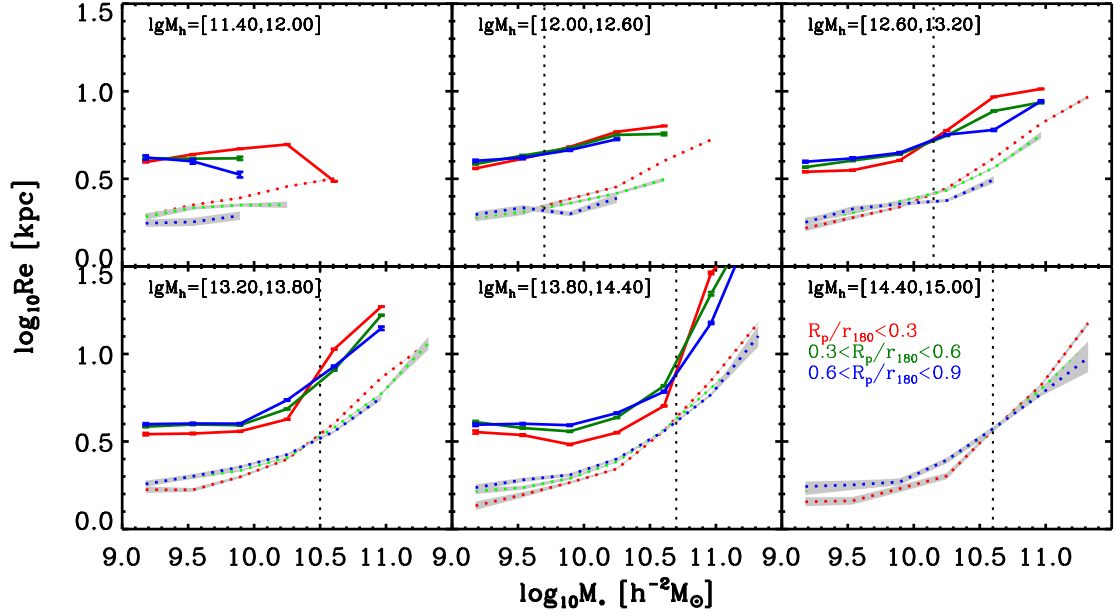,clip=true,width=0.90\textwidth} 
   \end{center}
  \caption{ The \mstar-\re\ relations for EAGLE galaxies with various
    halo mass and halo-centric radius as indicated in the panels. For
    comparison, the \mstar-\re\ relations of SDSS galaxies are shown
    in dotted lines, taken from the top group of panels of Figure
    \ref{fig:pos_mh_1th}. The observational results are multiplied by
    a factor of 1.4 to correct for the circular aperture effect (see
    text for details).  The vertical dashed lines are the same as
    those in Figure \ref{fig:pos_mh_1th}, indicating the transitional
    stellar mass of star formation quenching. }
  \label{fig:mstar_re_eagle_radius}
\end{figure*}

\begin{figure*}
   \begin{center}
    \epsfig{figure=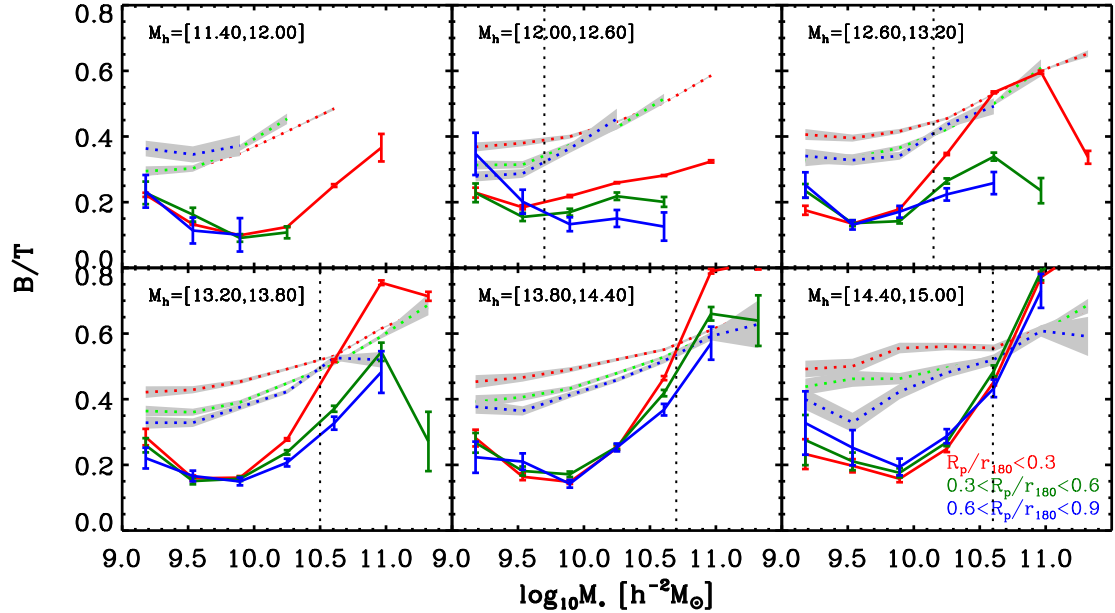,clip=true,width=0.90\textwidth} 
   \end{center}
  \caption{The \mstar-B/T relations for L-GALAXIES galaxies with
    various halo mass and halo-centric radius as indicated in the
    panels. For comparison, the results for SDSS galaxies are shown in
    dotted lines, taken from the bottom group panels of Figure
    \ref{fig:pos_mh_1th}. The vertical dashed lines are the same as
    those in Figure \ref{fig:pos_mh_1th}, indicating the transitional
    stellar mass of star formation quenching. }
  \label{fig:mstar_bt_lgals_radius}
\end{figure*}

We now focus on the dependence of the \mstar-\re\ and
\mstar-B/T relations on halo-centric radius, and investigate whether
current galaxy formation models can reproduce the transitional stellar
mass found in Section \ref{subsec:3.3}.  Figure
\ref{fig:mstar_re_eagle_radius} shows the \mstar-\re\ relation of
EAGLE galaxies in three bins of normalized halo-centric
distance, $R_{\rm p}/r_{\rm 180}$. For comparison, we also show the
results for SDSS galaxies as dotted lines.  Similar to Figure \ref{fig:mstar_eagle_1th}, the sizes of EAGLE
galaxies are overall $\sim$0.3 dex higher than those of SDSS galaxies,
almost over the entire stellar and halo mass
ranges. Interestingly, the dependence on $R_{\rm p}/r_{\rm
  180}$ for EAGLE galaxies is very similar to that for SDSS
galaxies: galaxies at smaller halo-centric distances are
smaller at the low-mass end and larger at the high-mass end. Most
remarkably, the corresponding transitional stellar masses for EAGLE
galaxies are almost identical to those for SDSS galaxies over the
entire halo mass range probed.

For a given halo mass bin, the difference in the mass-size relation between 
the two high halo-centric radius bins is insignificant, consistent with the result 
of SDSS galaxies shown in Figure \ref{fig:pos_mh_1th}. As in Section \ref{subsec:3.3}, 
we have also tried to detect a possible critical halo-centric radius 
in the EAGLE sample.  Such a critical radius was found, at least for halo masses 
above $10^{13.2}$ \msun $h^{-1}$, and its value is about $0.5 r_{\rm 180}$,
comparable to that observed for SDSS galaxies.

Figure \ref{fig:mstar_bt_lgals_radius} shows the B/T of L-GALAXIES
galaxies as a function of stellar mass at different $R_{\rm p}/r_{\rm
  180}$. For comparison, the \mstar-B/T relations for SDSS galaxies,
taken from Figure \ref{fig:pos_mh_1th}, are also plotted.  In addition
to the discrepancy in the overall trend between L-GALAXIES and SDSS
galaxies discussed in \S\ref{subsec:4.1}, the dependence of the
\mstar-B/T relation on $R_{\rm p}/r_{\rm 180}$ for L-GALAXIES is also
different from that of SDSS galaxies. The B/T of L-GALAXIES galaxies
is almost independent of $R_{\rm p}/r_{\rm 180}$ at low \mstar, and
decreases with increasing $R_{\rm p}/r_{\rm 180}$ at high \mstar,
which is opposite to the trend seen in the observational results.

\subsection{The effect of group finder algorithm}
\label{subsec:4.3}

The SDSS groups are identified using the halo-based group finder of
\citet{Yang-07}, which carries its own uncertainties in
central-satellite classification, group member identification and the
assignment of halo mass \citep[see e.g.][]{Campbell-15}.  In \citetalias{Wang-18b}
we therefore carried out a detailed investigation of how errors
  due to the group finder impact a comparison of centrals and
  satellites at given stellar and/or halo mass.  Here we examine
potential errors in the \mstar-\re\ (for EAGLE) and \mstar-B/T
(for L-GALAXIES) relations of centrals and satellites caused by
imperfections of the group finder. To this end, we apply the
group finder to the two mock galaxy catalogs and obtain two mock group
catalogs, to which we refer as L-GALAXIES+GF and EAGLE+GF,
respectively.  
For the two catalogs, halo masses are assigned to individual 
groups on the stellar mass of galaxies, consistent with the halo mass
estimate used for SDSS groups.

\begin{figure*}
  \begin{center}
    \epsfig{figure=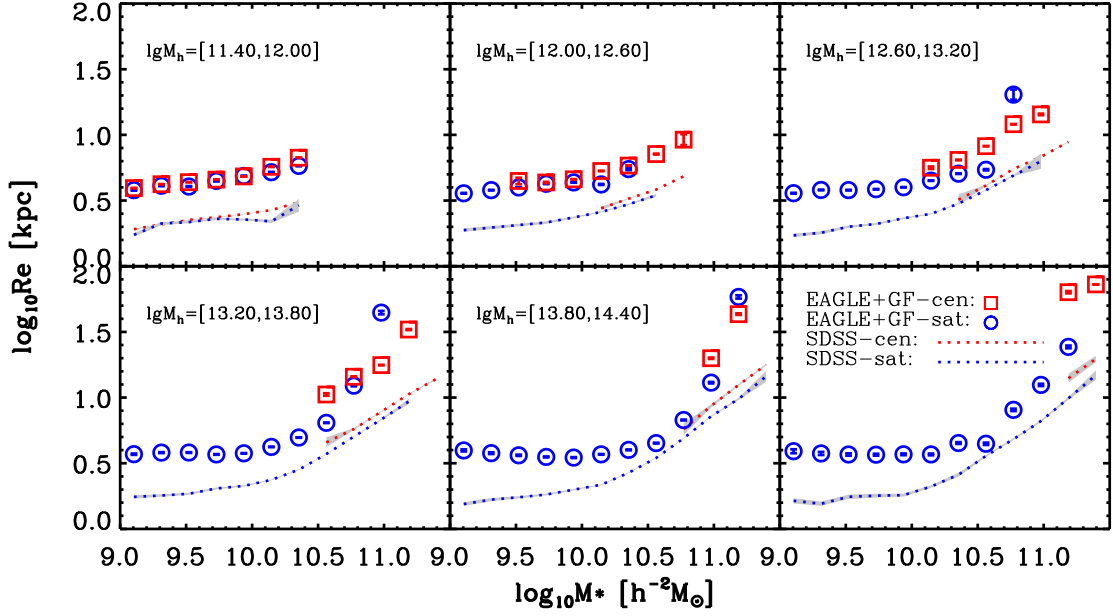,clip=true,width=0.90\textwidth} 
   \end{center}
  \caption{Similar to Figure \ref{fig:mstar_eagle_1th} but for
    EAGLE+GF catalog. It means that the central-satellite
    classification and halo masses for EAGLE galaxies are all given by
    the group finder algorithm. }
  \label{fig:mstar_eagle_2th}
\end{figure*}

\begin{figure*}
   \begin{center}
    \epsfig{figure=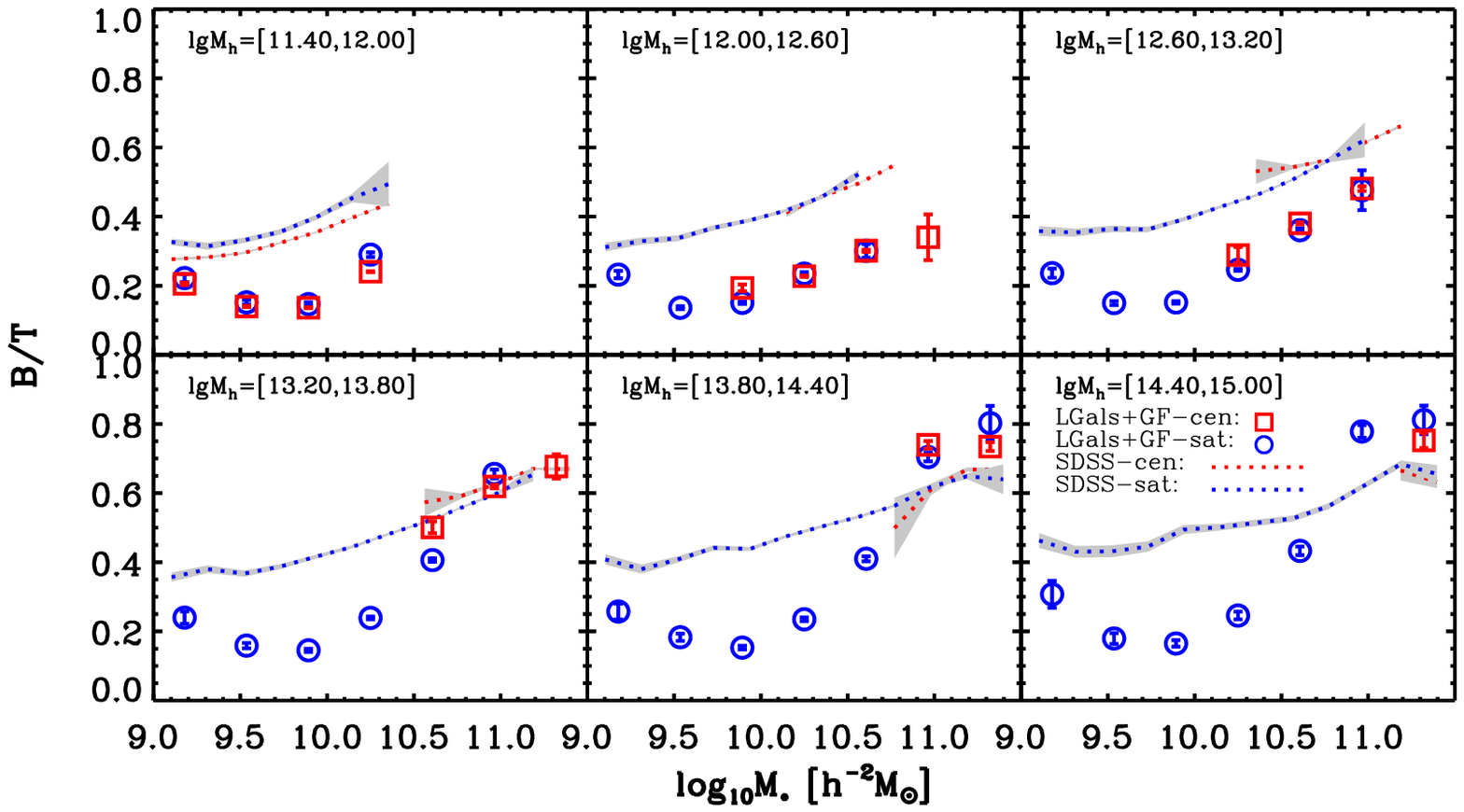,clip=true,width=0.90\textwidth} 
   \end{center}
  \caption{Similar to Figure \ref{fig:mstar_lgals_1th} but for
    L-GALAXIES+GF catalog. It means that the central-satellite
    classification and halo masses for L-GALAXIES galaxies are all
    given by the group finder algorithm.}
  \label{fig:mstar_lgals_2th}
\end{figure*}

\begin{figure*}
   \begin{center}
    \epsfig{figure=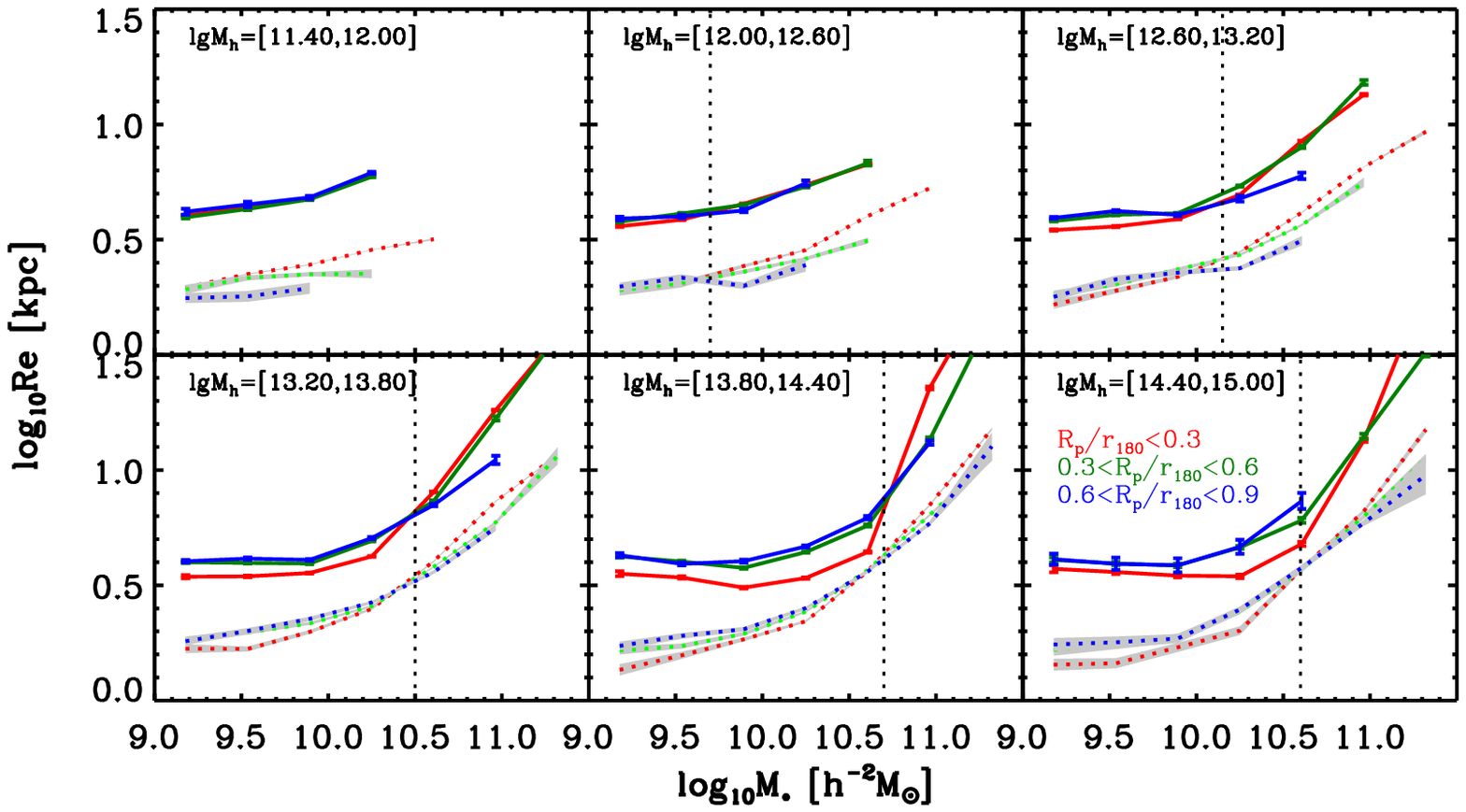,clip=true,width=0.90\textwidth} 
   \end{center}
  \caption{Similar to Figure \ref{fig:mstar_re_eagle_radius} but for
    EAGLE+GF catalog. The halo masses and central-satellite
    classification for EAGLE galaxies are given by the group finder
    algorithm.  }
  \label{fig:mstar_re_eagle_radius_2th}
\end{figure*}

\begin{figure*}
   \begin{center}
    \epsfig{figure=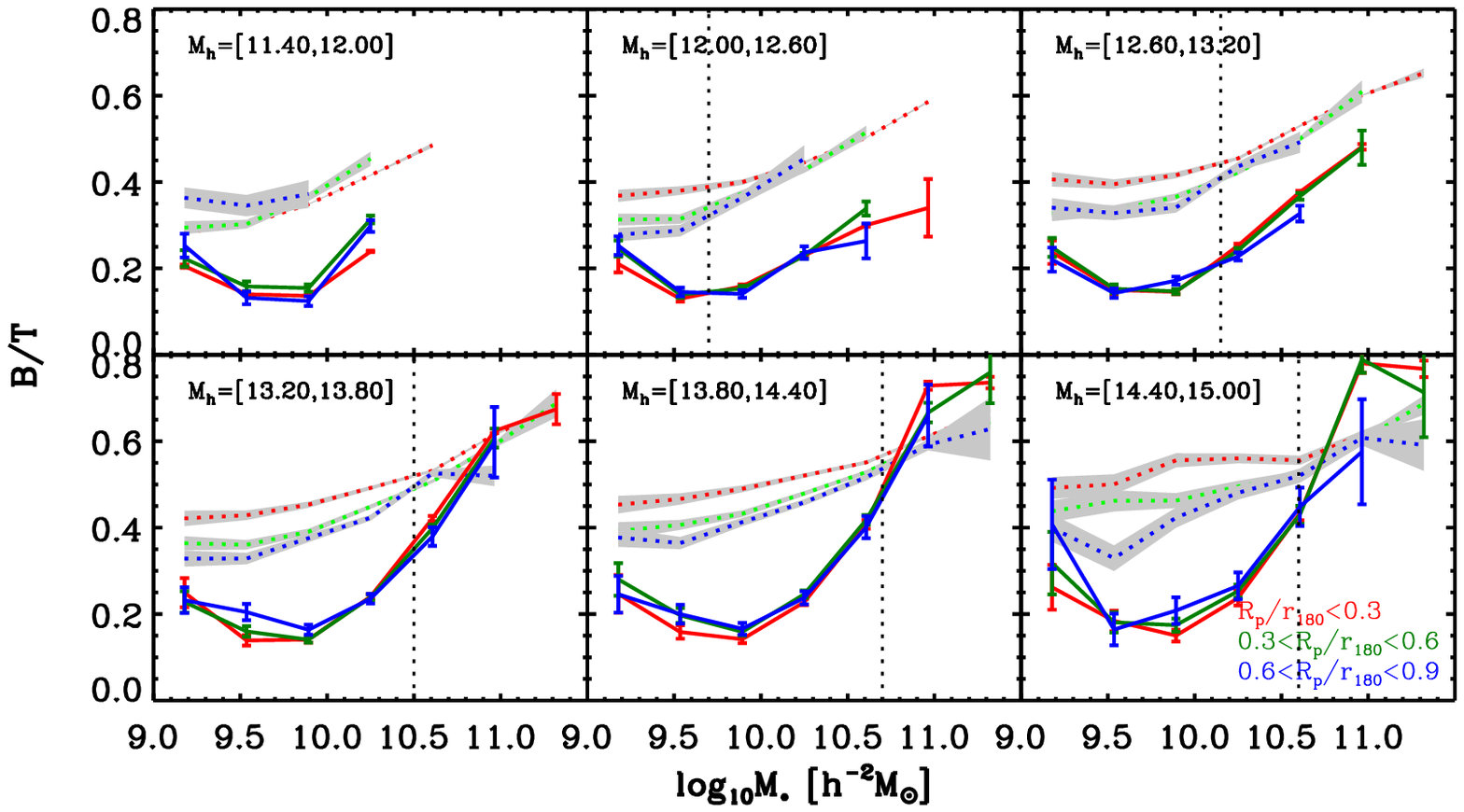,clip=true,width=0.90\textwidth}
   \end{center}
  \caption{Similar to Figure \ref{fig:mstar_bt_lgals_radius} but for
    L-GALAXIES+GF catalog. The central-satellite classification and
    halo masses for L-GALAXIES galaxies are all given by the group
    finder algorithm. }
  \label{fig:mstar_bt_lgals_radius_2th}
\end{figure*}

The results are shown in Figures \ref{fig:mstar_eagle_2th},
\ref{fig:mstar_lgals_2th}, \ref{fig:mstar_re_eagle_radius_2th} and
\ref{fig:mstar_bt_lgals_radius_2th}.  After applying the group finder,
the results for EAGLE change only very slightly. As
  before centrals and satellites reveal similar behavior (see Figure \ref{fig:mstar_eagle_1th}) and
the transitional stellar masses are apparent and very similar to
  those in Figure \ref{fig:mstar_re_eagle_radius}.
Furthermore, the overall trend in
the mass-size relation is broadly consistent with that for SDSS
galaxies. In contrast, in the case of L-GALAXIES, the
application of the group finder drastically reduces the
differences between centrals and satellites seen in Figure
\ref{fig:mstar_lgals_1th}. Moreover, the dependence of B/T on
halo-centric distance apparent in Figure
\ref{fig:mstar_bt_lgals_radius} also disappears, but the upturn in the
mass-B/T relation at the low-mass end remains (Figure
\ref{fig:mstar_bt_lgals_radius}).

Similar results were also obtained in \citetalias{Wang-18b}, where we
  demonstrated that the similarity in the quenched fraction between
centrals and satellites in EAGLE remains after applying the group
finder, while the significant differences in the quenched fraction
between centrals and satellites in L-GALAXIES was largely
reduced. This indicates that the performance of the group finder
depends on galaxy formation models. Indeed, further investigation
shows that the application of the group finder to EAGLE yields
more accurate halo mass assignments and better
central-satellite separations than in the case of L-GALAXIES
(see \citetalias{Wang-18b}). 
Specifically, for EAGLE, the group finder is able to 
reproduce the true halo mass with an overall offset of 0.03 dex and 
an overall scatter of 0.2 dex. For L-GALAXIES, on the other hand, 
the group finder reproduces the true halo mass with a larger offset, 
0.09 dex, and a larger scatter, 0.4 dex (see figure 8 in 
\citetalias{Wang-18b}). 

This leaves us with the somewhat uncomfortable situation that
  there are two possible explanations for the observed similarity
  between centrals and satellites in the SDSS: either there indeed is
  no intrinsic difference (at fixed stellar and halo mass), or the
  group finder has artificially erased an existing difference. And
  since it is unknown a priori whether L-GALAXIES or EAGLE is a better
  representation of the real universe, it is unclear which
  interpretation is more appropriate for SDSS galaxies. In \citetalias{Wang-18b},
we addressed this conundrum by proposing an independent test to
check the performance of the group finder.  The test calculates the
clustering of galaxy groups identified by the group finder as a
  function of their assigned halo mass and examines whether or not
the resulting halo bias is consistent with theoretical
predictions.  If the group finder performs poorly, the assigned
halo masses and member identifications should have large
errors, causing the inferred `halo-bias' to deviate from
theoretical expectations. This test can be applied not only to galaxy
formation models such as L-GALAXIES and EAGLE, but also to the
actual SDSS data. Our examination in \citetalias{Wang-18b} clearly shows
that the resulting halo bias for EAGLE+GF agrees well with the
prediction of \cite{Sheth-Mo-Tormen-01}, while the halo bias
  inferred from L-GALAXIES+GF differs strongly from these theoretical
  expectation. This is consistent with the direct examination
of halo mass assignments: for L-GALAXIES, the halo masses assigned by
the group finder carry large errors;
in the case of EAGLE the errors are significantly smaller. 
Applying the `halo-bias' test
  to the SDSS group catalog, we obtain results that are again in good
  agreement with the theoretical prediction
  \citep[][]{Wang.etal.08}. These results suggest that EAGLE is
  likely to be more reminiscent of the real universe and that the
group finder is expected to work well for SDSS galaxies. And
  although not a watertight argument, this does suggest that centrals
  and satellites really are very similar at fixed stellar and halo
  mass.

When applying the group finder to EAGLE and L-GALAXIES, we correct
the spectroscopic in-completeness due to the magnitude limit by 
using the observed luminosity function to account for the contribution 
of the missing galaxies. However, we do not include the in-completeness due 
to the fiber collisions\footnote{The in-completeness due to fiber collisions
can have a significant effect on
the richness of individual groups, but does not have a significant impact
on the halo mass assignment \citep{Yang-07}.} and the photometric uncertainties
(or stellar mass uncertainties).
Including these would undoubtedly make the performance of group finder worse. 
We have examined our results by using 
halo mass assignment based on the $r$-band luminosity for EAGLE+GF, and found 
that the results shown in Figures \ref{fig:mstar_eagle_2th} and 
\ref{fig:mstar_re_eagle_radius_2th} are not affected. This suggests that 
the uncertainties in converting $r$-band luminosity to stellar mass 
should not change our results. This is consistent with the finding 
in \cite{Yang-07} that halo masses based on the $r$-band luminosity 
and the stellar mass agree with each other (see their figure 10).

\section{Summary and Discussion}
\label{sec:summary}

In \citetalias{Wang-18a}, we found that centrals and satellites essentially have the
same quenched fraction as long as both the stellar and halo masses are
controlled. The prevalence of optical-selected/radio-selected AGN is
also found to be similar for centrals and satellites at given stellar
masses. In the present paper, as an extension of \citetalias{Wang-18a}, we
investigate the sizes and bulge-to-total light ratios of centrals and
satellites to analyze to what extent ``central versus satellite''
impacts the size and structural properties of galaxies. In particular,
we use the SDSS galaxy groups from \cite{Yang-07} to examine the size
and bulge-to-total light ratio as functions of galaxy stellar mass,
and how the relations are affected by the host halo mass and
halo-centric distance.  We also compare the observational results with
two galaxy formation models, the latest version of a semi-analytic
model, L-GALAXIES, and the state-of-the-art hydrodynamical simulation,
EAGLE.  Our main results can be summarized as follows.

\begin{itemize}
 
\item At a given stellar mass, central galaxies have larger
  sizes and smaller bulge-to-total ratios than satellites.
  However, when galaxies are separated into a set of narrow halo mass
  bins, the differences between centrals and satellites
  disappear. Thus, centrals and satellites have similar mass-size and
  mass-B/T relations as long as halo masses are controlled.
  
\item When the stellar mass distribution is controlled
to be the same, centrals and satellites in halos of similar mass 
have similar size and B/T distributions.

\item The dependences of size and B/T on halo mass become weaker as
  stellar mass increases.  For galaxies with stellar masses
  $>10^{10.7} h^{-2}$\msun, neither size nor bulge-to-total
    ratio shows any dependence on halo mass at fixed stellar mass. At
    \mstar $<10^{10.7} h^{-2}$\msun, sizes decrease and B/T increase
    with increasing halo mass.

\item The mass-size and mass-B/T relations as functions of the
  normalized halo-centric distance show a transitional stellar mass,
  below which the galaxy size (B/T) increases (decreases) with
  increasing halo-centric distance, while above which the dependence
  is reversed for size and disappears for B/T.

\item The transitional stellar masses for the size and B/T are similar
  to each other and to that seen in the quenched fraction, suggesting
  that star formation quenching, size and morphology evolution may all
  have related origins. The transitional stellar mass is about
  one-fifth of the stellar mass of centrals in the host halos.

\item The EAGLE simulation successfully reproduces the similarity of
  the mass-size relation for centrals and satellites at given halo
  masses, although the overall size is about 0.3 dex higher than that
  of SDSS galaxies.  In contrast, L-GALAXIES fails to reproduce the
  similarity of the mass-B/T relation for centrals and satellites.

\item The EAGLE simulation successfully reproduces the observed
  dependence of galaxy size on halo-centric distance. The transitional
  stellar masses predicted by EAGLE also match well those obtained for
  SDSS galaxies. The L-GALAXIES fails to reproduce the observed
  dependence of B/T on halo-centric distance.

\end{itemize}


Combining these results with those from \citetalias{Wang-18a}, we find that
  centrals have similar quenched fractions, a similar prevalence of
  radio/optical-selected AGN, and similar sizes and structural
  properties as satellites of similar stellar mass, as long as the
  halo masses are constrained to a narrow range.  This indicates that
  the differences between centrals and satellites found in numerous
  previous investigations \citep[e.g.][]{vdBosch-08, Pasquali-09,
    Peng-12, Bluck-16, Spindler-Wake-17}, mainly reflects that they
  were not controlled for stellar and/or halo mass.
  
  Based on these findings, two different explanations emerge to
  account for the observational results. One natural scenario is 
  that ``being a central'' is not special compared to ``being a
  satellite''; their main difference is that they occupy different
  regions of the stellar mass-halo mass (or stellar
  mass-environment) parameter space.  Specifically,  the
  environmental processes, such as tidal stripping, ram-pressure
  stripping and harassment, work in the same way to galaxies of
  similar stellar mass in the same position of groups or clusters, 
  regardless of ``being a central'' or ``being a satellite''.  However,
  it is also possible that the environmental processes indeed work
  on centrals and satellites in different ways, while the more
  massive satellites have been accreted more recently and have no
  sufficient time to be significantly affected by
  “satellite-specific” processes.  Note that the two scenarios are
  not necessarily in conflict with each other; both may happen in
  reality. We discuss the above two scenarios in details
  respectively below.

The former scenario indicates a stellar mass-dependent environmental
processes\footnote{Because centrals and satellites are believed to
be the same under this scenario,  we do not call some of the environmental
processes as the ``satellite-specific'' processes. However, in the second
scenario, the ``satellite-specific'' processes that work only on
satellites are assumed  to be exist.}, suggesting that a more
fundamental parameter related to the environmental effects is the
stellar mass of the galaxies at a given position of a given group,
rather than “being a central” or “being a satellite”. Indeed, evidences
have been found to reduce the difference between central and massive
satellites. For instance, in the observations and hydrodynamical
simulations, the hot gas reservoirs of massive satellite galaxies can
survive for some considerable time after falling into a cluster
\citep{Sun-07, Jeltema-08, Weinmann-10}.  
\cite{McCarthy-08} found that satellites with typical
structural and orbital parameters can maintain up to 30 per cent of 
the initial hot halo gas for up to 10 Gyr.  Furthermore,  using
hydrodynamic simulation, \cite{Keres-09} found that both centrals 
and satellites of similar baryonic mass originally acquire most of 
their baryonic mass through filamentary “cold mode” accretion of gas
with similar gas accretion rates at z$>$1.  These evidences indicate 
that massive satellites can acquire cold gas via both cooling of
surrounding hot gas reservoirs and direct filamentary ``cold
mode'' accretion, which resemble the properties of centrals. This may 
be the reason for the similar quenching properties of centrals and
satellites when both stellar mass and halo mass are controlled.  

Thus, the simple scenario that the effect of environmental processes only depend on the stellar mass of galaxies and their location in groups at given halo mass, appears to be able to explain the similarity of centrals and satellites found in \citetalias{Wang-18a} and this work.  This scenario is not unreasonable, because the effects of some of the environmental processes are indeed related to the gravitational potential and/or the structure of the galaxies.  For instance, more massive galaxies are not easily disturbed in galaxy-galaxy interaction and/or harassment. Ram pressure stripping is more inefficient for galaxies with higher stellar mass surface density, usually corresponding to more massive galaxies. 

The later scenario is that the massive satellites are newly accreted
and have not yet been affected by the ``satellite-specific'' processes. 
In this case, the origin of the transitional stellar mass
is more related to the time since accretion of the satellite galaxies.
More massive satellite galaxies are likely to have been accreted more
recently, for three reasons; first of all, massive satellites
experience strong dynamical friction, and therefore have a limiting
`lifetime' before being cannibalized. Secondly, the fact that the
satellites are massive, with a stellar mass close to that of the
central, means that they cannot (yet) have experienced much tidal
stripping. And finally, massive satellites reside in massive subhalos,
and, due to the hierarchical nature of structure formation, more
massive subhalos are typically accreted later \citep[][]{vdBosch-16}.
Hence, the fact that satellites with a stellar mass above the
transitional mass look similar to centrals, with no obvious sign of
having been affected by satellite-specific processes, may simply
indicate that said processes have not yet had sufficient time to
significantly affect the star formation rate or structural properties
of the satellites in question. Indeed, tidal and ram-pressure
stripping are unlikely to affect the stellar mass of the satellite
galaxy until most of the halo has been stripped down to the extent of
the stellar body, which can take multiple pericentric
passages. Strangulation, i.e., choking off the supply of new gas onto
the subhalo, only affects the star formation rate of a satellite
galaxy once it has run out of its cold gas supply, which can also take
several Gyrs. Within this picture, massive satellite simply look
identical to centrals simply because they were centrals until fairly
recent.

Although both of the suggested scenarios can explain the similarity 
of centrals and satellites found in \citetalias{Wang-18a} and this work, 
it is unclear which one is more likely to be real.  
Fortunately, the fact that galaxies with \mstar $> M_{\rm *,t}$ 
do show a dependence of size on halo-centric radius may provide
clues.  Actually, the dependence of galaxy size on environment 
have been found in decades, and minor merger has been proposed 
to explain the observational result that galaxies in dense 
environment are larger than their counterparts in less dense 
environment \citep[e.g.][]{Cooper-12, Strazzullo-13, Delaye-14,
 Yoon-Im-Kim-17, Huang-18}.  This does not seem to support 
 the second scenario that the massive satellites are accreted 
 more recently.  

Neglecting which scenario is more likely to be real,
an interesting implication of our results is that 
the commonly adopted `split' in centrals and satellite galaxies, 
which is often used to assess the importance of environment on 
galaxy formation and evolution may not be optimal.
In this work, we find that galaxies below the transitional stellar 
mass, roughly one-fifth of that of the (average) central, 
have quenched fractions, sizes and bulge-to-total ratios that depend
on halo-centric radius, indicating that they are experiencing
environmentally-induced changes to their properties. Above the
transitional stellar mass, galaxy properties show virtually no
dependence on halo-centric distance\footnote{An exception is galaxy
  size, which becomes larger at smaller halo-centric distances; this
  trend is opposite to that for galaxies below the transitional
  stellar mass.}.
According to this, we argue instead that splitting the galaxy
population based on the transitional stellar mass $M_{\rm *,t}(M_{\rm
  h})$ identified here might be more fundamental, in that it more
accurately separates those galaxies that have been affected by
satellite-specific processes from those that have not.

An intriguing aspect of the transitional stellar mass is that it is
basically the same for quenching properties, galaxy sizes, and
structural properties such as the dominance of the bulge.  There is no
shortage of suggested explanations for this. For instance, it has been
suggested that the quenching of star formation is directly associated
with a morphological transformation \citep[e.g.][]{Fujita-98, Read-06,
  Brennan-15, Pawlik-18}, or with the `apparent' evolution in galaxy
size due to fading of the disk \citep[e.g.,][]{Weinmann-09}. Others
have suggested that quenching preferentially occurs in more compact
galaxies \citep[e.g.][]{Fang-13, Barro-17, Wang-Kong-Pan-18,
  Socolovsky-19, Wang-19a}, possibly due a shorter gas depletion time
and/or enhanced feedback from more massive black holes in more compact
star forming galaxies. A subtly different idea is that quenched
galaxies have formed earlier, when the Universe was denser resulting
in more compact galaxies \citep[e.g.][]{Carollo-13a, Daz-Garca-19}.
Finally, \cite{Lilly-Carollo-16} have pointed out that the close link
between quenching and evolution in structural properties does not
necessarily imply a physical causation; it can simply arise from a
combination of a mass-dependent quenching law and the observed
evolution of the size-mass relation for star-forming galaxies.

To fully uncover the origin of the transitional stellar mass, 
there is an obvious path forward. As we have
demonstrated, the hydrodynamical EAGLE simulation accurately
reproduces the trends identified in this paper. In particular, it
reveals the same transitional stellar mass as function of halo mass as
in the SDSS data, and even reveals the same, opposite trends between
halo-centric radius and galaxy size above and below this transitional
stellar mass.  A detailed study of the evolution of galaxies in the
EAGLE simulation therefore should be able to shed important light on
how galaxies quench, and evolve in size and bulge-to-total ratio.
Contrasting this with what happens in the semi-analytical model
L-GALAXIES, which fails to reproduce many of the trends presented
here, will also help in furthering our understanding of the intricate
processes that underly galaxy formation and evolution. In a
forthcoming paper we will follow this approach and analyze the
detailed evolution of galaxies in both the EAGLE simulations and the
L-GALAXIES semi-analytical model. Finally, we also advocate comparing
the SDSS results presented here and in \citetalias{Wang-18a} with other models for galaxy
formation, including state-of-the-art hydrodynamical simulations such
as Illustris-TNG \citep[][]{Pillepich-18} and Horizon-AGN
\citep[][]{Dubois-14}, semi-analytical models such as GALFORM
\citep[][]{Cole-00} and Galacticus \citep[][]{Benson-12}, and even
empirical models such as EMERGE \citep[][]{Moster-18} and the
UniverseMachine \citep[][]{Behroozi-19}.

\acknowledgments

We thank the anonymous referee for constructive and thoughtful
suggestions to improve the paper. 
This work is supported by the National Key R\&D Program of China
(grant No. 2018YFA0404503) , the National Basic Research Program of
China (973 Program)(2015CB857002), the National Natural Science
Foundation of China (NSFC, Nos.  11733004, 11421303, 11890693,
11673015, 11522324, 11433005, 11320101002, 11833005, 11890692, and
11621303), and the Fundamental Research Funds for the Central
Universities. 
EW is supported by the Swiss National Science Foundation. 
FvdB is supported by the US National Science
Foundation through grant AST 1516962, by the National Aeronautics
 and Space Administration through Grant No. 17-ATP17-0028, and by the Klaus Tschira foundation.
The work is also supported by the Supercomputer Center
of University of Science and Technology of China.

\bibliography{rewritebib_Frank.bib}

\appendix

\section{The examination of the stellar surface density within \re\ for centrals and satellites.}

\begin{figure*}
  \begin{center}
      \epsfig{figure=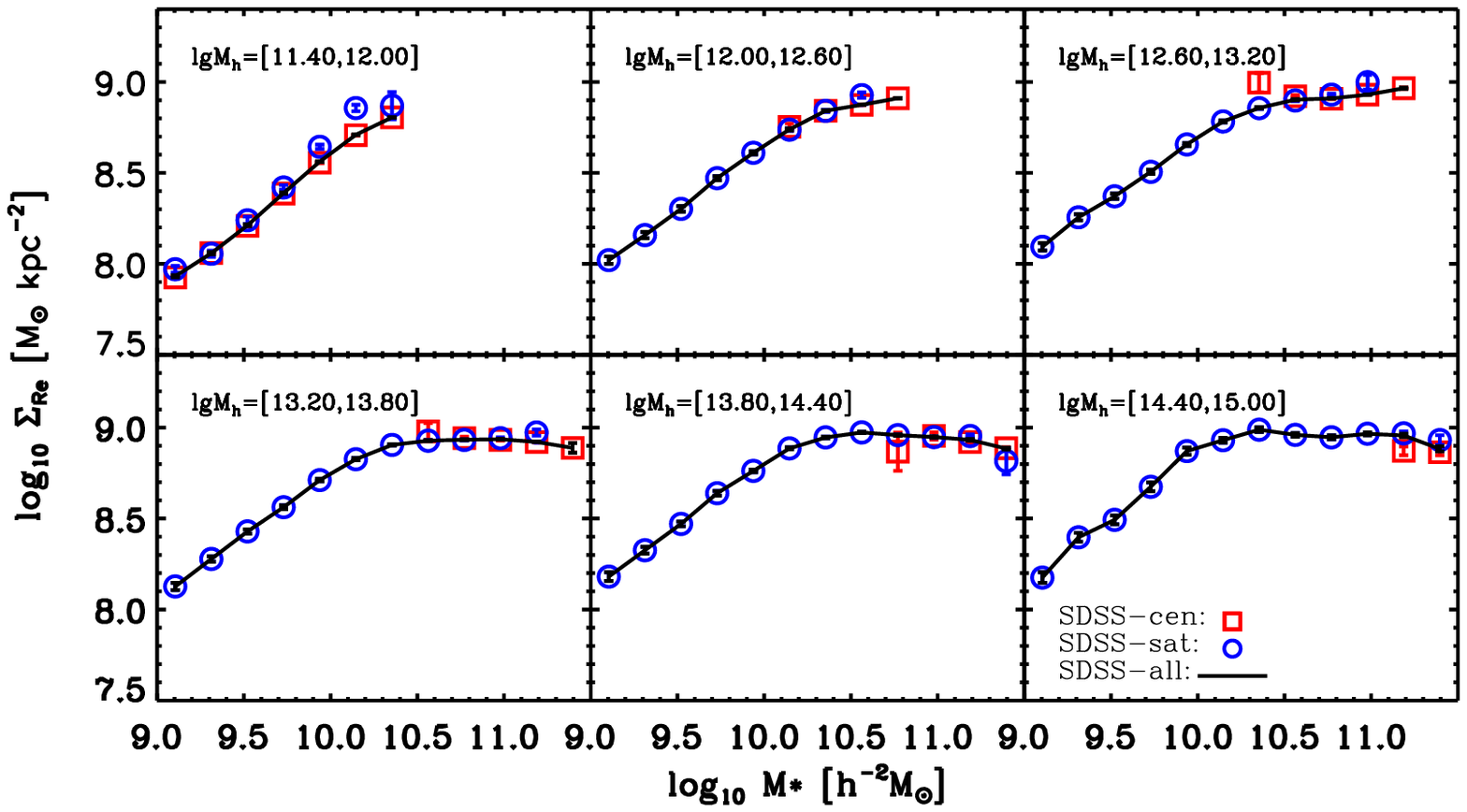,clip=true,width=0.8\textwidth}
      \epsfig{figure=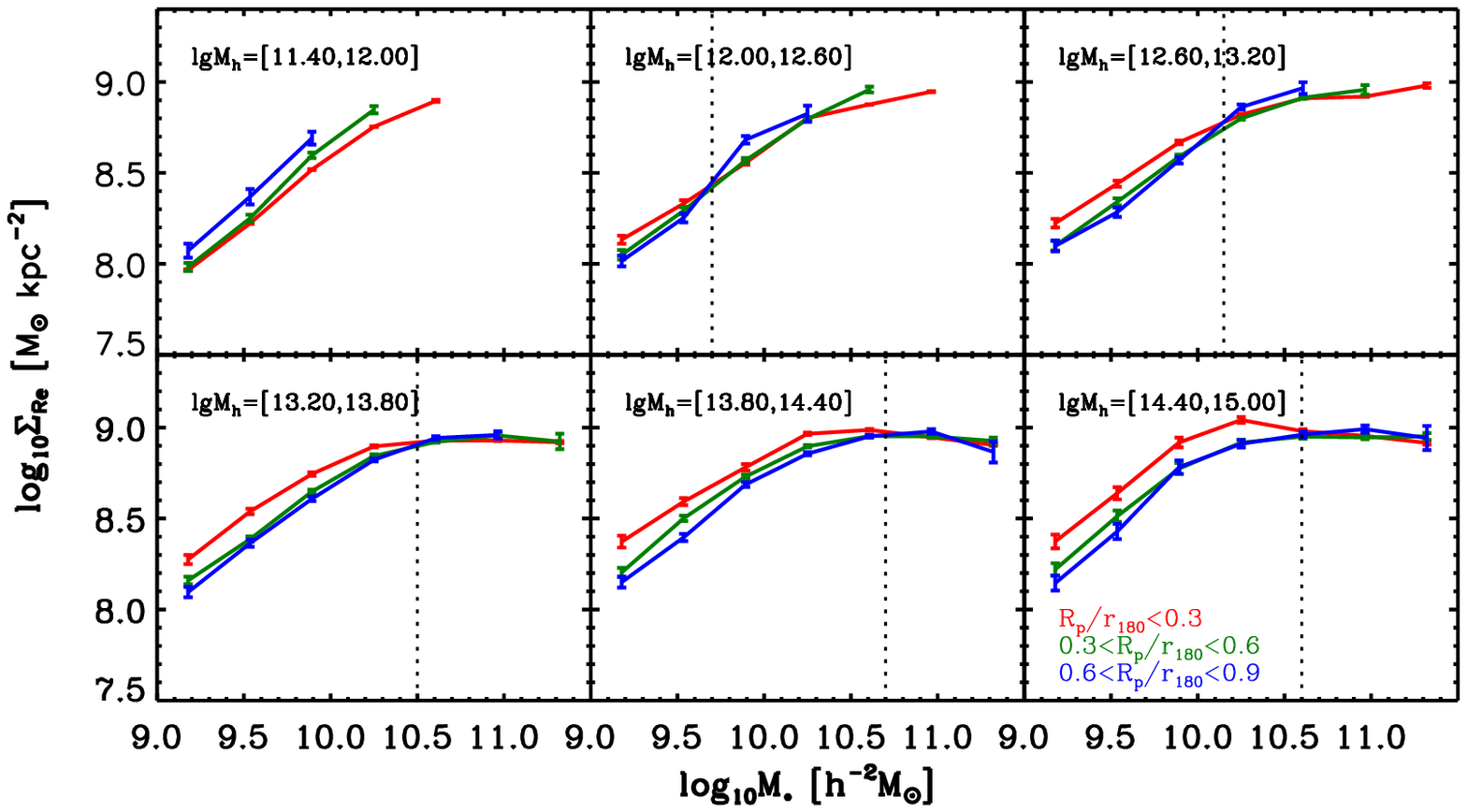,clip=true,width=0.8\textwidth}
  \end{center}
  \caption{Top group of panels: The \mstar-$\Sigma_{\rm Re}$ relations for centrals 
  and satellites at a set of halo mass bins.   In each panel,
  the results for centrals, satellites and all galaxies are
  indicated by red squares, blue circles and black line,
  respectively.
  Bottom group panels: The \mstar-$\Sigma_{\rm Re}$ 
  relations for galaxies at three different halo-centric 
  bins, in the same set of halo mass bins.  }
  \label{fig:mure_mass_1th}
\end{figure*}

In the main text, we have compared the half-light radius and the
bulge-to-light ratio of centrals and satellites with both
stellar mass and halo mass controlled. We found that centrals and satellites 
in halos of a given mass have
similar mass-size and mass-B/T relations. 
However, the measurement of B/T may depend on the model of the
bulge assumed in the decomposition.  Here, we present the 
comparison between centrals and satellites in the same way as in 
Sections \ref{subsec:3.2} and \ref{subsec:3.3}, but using stellar 
surface density within the effective radius, $\Sigma_{\rm Re}$. 
This parameter can be estimated without an assumed model for 
the light profile of a galaxy. 

The $\Sigma_{\rm Re}$ is calculated by directly integrating the light
profiles from the innermost point out to \re, adopting the relation
between $M_*/L_{i}$ (ratio between stellar mass and $i$-band
luminosity) and the rest-frame $g-i$ color from \cite{Fang-13}: 
$\log_{10} M_*/L_{i}=1.15+0.79\times(g-i)$. In practice, we generate
the cumulative flux profile at a series of radii and obtain the total
flux within \re\ by the cubic spline interpolation for the $g$ and $r$ 
bands, based on the azimuthally averaged radial surface brightness profile 
(also known as {\tt ProfMean} in the output of the SDSS pipeline). 
The $i$-band luminosity and $g-i$ color are corrected to the rest frame
\cite{Blanton-Roweis-07} and for Galactic extinction
\citep{Schlegel-98}. The {\tt ProfMean} 
provided by the SDSS pipeline is based on the light 
profile in circular aperture. To obtain the $\Sigma_{\rm Re}$ 
in a consistent way, we adopt the 
\re\ from the NYU-VAGC catalog \citep{Blanton-05a}, which is 
measured also with circular aperture.

The top group panel of Figure \ref{fig:mure_mass_1th} shows the
mean $M_*-\Sigma_{\rm Re}$ relation
for centrals and satellites at the six halo mass bins. 
As one can see, when both halo mass and stellar mass are 
controlled, centrals and satellites show similar 
$M_*$-dependence of $\Sigma_{\rm Re}$.  
The bottom group panel of Figure \ref{fig:mure_mass_1th} shows 
the $M_*-\Sigma_{\rm Re}$ relation for galaxies (including 
both centrals and satellites) in three bins of halo-centric radius
and six bins of halo mass. The results here are similar 
to those shown in Figure \ref{fig:pos_mh_1th}. 
The $M_*-\Sigma_{\rm Re}$ relation depends on 
$R_{\rm p}/r_{\rm 180}$ at stellar mass below the transitional 
stellar mass, while this dependence disappears for more massive 
galaxies. All these results are in good agreement with 
those presented in Section \ref{sec:results}.

\label{lastpage}
\end{document}